\documentclass[%
reprint,
superscriptaddress,
 amsmath,amssymb,
 aps,
prb,
floatfix]{revtex4-2}
\usepackage{libertine}
\usepackage{graphicx}
\usepackage{dcolumn}
\usepackage{bm}
\usepackage{hyperref}
\usepackage{upgreek}

\begin{document}


\title{\sc \Large PyMoosh : a comprehensive numerical toolkit for computing the optical properties of multilayered structures}
\author{Denis Langevin}
\email{denis.langevin@uca.fr}
\affiliation{Universit\'e Clermont Auvergne, Clermont Auvergne INP, CNRS, Institut Pascal, F-63000 Clermont-Ferrand, France}
\author{Pauline Bennet}
\affiliation{Institut Matériaux Microélectronique Nanosciences de Provence (IM2NP),\\ CNRS, Aix-Marseille Université, Marseille, France}
\author{Abdourahman Khaireh-Walieh}
\author{Peter Wiecha}
\affiliation{LAAS, Universit\'e de Toulouse, CNRS, Toulouse, France}
\author{Olivier Teytaud}
\affiliation{Meta AI Research Paris, France}
\author{Antoine Moreau}
\email{antoine.moreau@uca.fr}
\affiliation{Universit\'e Clermont Auvergne, Clermont Auvergne INP, CNRS, Institut Pascal, F-63000 Clermont-Ferrand, France}

\date{\today}

\begin{abstract}
We present  PyMoosh, a Python-based simulation library designed to provide  a comprehensive set of numerical tools allowing computation of essentially all  optical characteristics of multilayered structures, ranging from reflectance and transmittance to guided modes and photovoltaic efficiency. PyMoosh is designed not just for research purposes, but also for use cases in education. To this end, we have invested significant effort in ensuring user-friendliness and simplicity of the interface.  PyMoosh has been developed in line with the principles of Open Science and taking into account the fact that multilayered structures are increasingly being used as a testing ground for optimization and deep learning approaches. In this paper, we provide the theoretical basis at the core of PyMoosh, an overview of its capabilities, and a comparison between the different numerical methods implemented in terms of speed and stability. We are convinced that such a versatile tool will be useful to the community in many ways.
\end{abstract}

\maketitle

\section{Introduction}

In optics, multilayered structures have been studied extensively for more than a century. The very first effort at calculating the reflectance of multilayers dates back to the early work of Lord Rayleigh\cite{strutt1912propagation,strutt1917reflection}.
The subject gained increasing attention in the middle of the twentieth century when quarter-wave stacks (also called Bragg mirrors or 1D photonic crystals) were fabricated and fully understood for the first time\cite{macleod2012quarterwave}. An important contribution has been the formalism introduced by Abélès\cite{Abeles1950theorie}, which provided a deeper insight into the physical mechanisms behind the optical properties of multilayer systems, as well as a convenient mathematical framework for their calculation.In the years that followed, this formal approach was extended and successfully applied to periodic structures\cite{yeh1977electromagnetic,yariv1977electromagnetic,yeh1978optical} and anisotropic media\cite{yeh1980optics}. The resulting formalisms have been widely incorporated in optical textbooks\cite{born2013principles,yeh2006optical,macleod2017thin}.

The design of diverse optical filters, guided by physical intuition, emerged as early as 1958\cite{baumeister_design_1958}, and thrived in the subsequent years, notably due to the works of Thelen\cite{thelen_equivalent_1966,thelen_design_1971,apfel_optical_1977}. The first computer program that allowed forowed for the automation of optical filter design was published by Dobrowolsky and Lowe in 1978\cite{dobrowolski_optical_1978}.
Given the versatility of multilayered structures, very efficient design methods have been conceived since then \cite{tikhonravov1994development}, requiring the reliable computation of the optical properties of increasingly complex filters. 
A direct application has been the Wavelength-Division Multiplexing technology, which allowed a substantial increase of the transcontinental optical connection bandwidth in recent decades. 
Today, optical filter design and fabrication contests\cite{poitras_manufacturing_2017,kruschwitz_optical_2017,Kruschwitz:19} are periodically organized, showing how powerful state-of-the-art techniques have become. In the late 20th century, multilayered structures gained renewed interest and became widespread due to advances in fabrication techniques. In particular, two types of problems have received particular attention due to their practical importance: sensors based on surface plasmon resonances for the detection of biologically relevant molecules\cite{bockova2019advances}, and anti-reflective coatings used in photovoltaic devices\cite{raut2011anti}. 
In parallel, the increasingly complex problems associated with plasmonics\cite{bozhevolnyi2007general}, metamaterials\cite{shekhar2014hyperbolic} or non-specular phenomena\cite{polles2016leveraging} have also made multilayer simulations more challenging, particularly due to the frequent need to include a large number of metallic layers in the structures.

Over the years, an increasing number of numerical tools have thus been proposed to compute the optical response of complex multilayers\cite{katsidisGeneralTransfermatrixMethod2002, luceTMMFastTransferMatrix2022, bayPyLlamaStableVersatile2022}, with a significant portion of them tailored for specific applications such as optical filters\cite{larouche2008openfilters} or surface plasmon resonance (SPR)\cite{costa2019sim}. The demand for accurate and efficient computation of properties has grown significantly, while well-studied and understood structures serve as a testing ground for optimization techniques \cite{barry2020evolutionary,wankerl2022directional}.
In the recent past, thanks to the fast computation methods, multi-layer stacks have also often been used as large and cost-effective datasets for rapid testing of deep learning algorithms 
\cite{liuGenerativeModelInverse2018, unniDeepConvolutionalMixture2020, daiAccurateInverseDesign2021, daiInverseDesignStructural2022, wangEllipsoNetDeeplearningenabledOptical2022, luceInvestigationInverseDesign2023, maOptoGPTFoundationModel2023}.
Due to its relative simplicity, differentiable transfer matrix implementations have furthermore been used as actual part of deep learning models for accelerated training \cite{jiangMultiobjectiveCategoricalGlobal2020, shiyuanliuSmartEllipsometryPhysicsinformed2023}.
Because of this multiplication of objectives, the range of computed quantities of physical interest has also expanded.

Here we introduce PyMoosh (Python-based Multilayer Optics Optimization and Simulation Hub), a user-friendly and comprehensive set of numerical tools to compute as efficiently and as reliably as possible many optical properties of multilayered structures\cite{moreau_pymoosh_2023, pymoosh}. 
This includes reflectance and transmittance in any condition and with different state-of-the-art numerical methods, but also absorption and Poynting flux in any layer, 2D Green functions, and the computation of structure modes and their dispersion curves. Finally, PyMoosh is specifically designed to be used together with optimization tools. Developed in Python, PyMoosh is an open-source software aimed at maximizing usability. It is intended to be simple enough to be used for educational purposes, while also offering advanced features for research, with a particular focus on tackling demanding problems such as optimization and deep learning. It is the successor of Moosh, which was implemented in Octave/Matlab \cite{defranceMooshNumericalSwiss2016}.

This article is structured as follows. In a first section, we detail the different techniques included in PyMoosh to solve Maxwell's equations in multilayered structures. PyMoosh specifically includes transfer and scattering matrices, the Abélès formalism, Dirichlet-to-Neumann maps and the admittance formalism.
The methods, all gathered and described in a single place for educational and convenience purposes, are compared in terms of computational costs and tested for reliability on common photonics problems. 
In a second section we detail how the Poynting flux and absorption in each layer can be computed and explain the tools specific to photovoltaic problems that are implemented in PyMoosh.
In a third section we explain how the field inside the structure can be computed with high accuracy using scattering matrices.
We demonstrate this for two example illuminations, namely an incoming beam and an oscillating line current source, placed at an arbitrary location inside the structure.
The fourth section describes how resonant modes (leaky or guided) can be retrieved.
The last section describes our general approach in terms of coding and implementation, along with  a discussion of the motivations behind our choices. PyMoosh comes in fact with a very large number of examples, under the form of Jupyter Notebooks\cite{jupyter}, which are used to illustrate its capabilities, and that may serve as starting points for applications of other users. 



\section{Computing reflectance and transmittance}

Historically, and for understandable practical reasons, the focus of mulilayer structure studies has been put on computing the reflectance and eventually the transmittance of a given structure for a given wavelength $\lambda$. This is equivalent to solving Maxwell's equations in the structure for a time dependency $e^{-i\omega t}$ where $\omega = \frac{2 \pi c}{\lambda}$ is the angular frequency of the electromagnetic wave.
Since the structure is invariant by translation along the $x$ and $y$ axis, we choose without any loss of generality, to consider a $e^{i k_x x}$ spatial dependency with respect to $x$ and invariance along $y$
for all fields. In Transverse Electric (TE) (resp. Transverse Magnetic (TM)) polarization, $E_y$ (resp. $H_y$) satisfies the following Helmholtz equation:
\begin{equation}
\partial_z^2 E_y + \left(\frac{\mu_r\epsilon_r \omega^2}{c^2} - k_{x}^2\right) E_y = 0\,,\label{eq:Helmholtz}
\end{equation}
in any linear medium characterized by its relative permittivity $\epsilon_r$ and permeability $\mu_r$. While the contribution of $\mu_r$ is often neglected because natural materials do not involve any magnetic response at optical wavelengths ($\mu_r \approx 1$ due to the small size of the Bohr radius in atoms\cite{giessenGlimpsingWeakMagnetic2009}), metamaterials can be tailored to form materials with an effective magnetic response.

Because the $k_x$ spatial dependency is constant through the stack, in any layer $i$, the vertical ($z$) propagation constant of the field can be defined by:
\begin{equation}
    \gamma_i= \sqrt{\mu_i\,\epsilon_i \,k_0^2-k_x^2}\,,
    \label{eq:gamma}
\end{equation}
where $k_0=\frac{\omega}{c} = \frac{2\pi}{\lambda}$ and where we write the relative permeabilities and permittivities of the $i$-th layer's material as $\mu_i$ and $\epsilon_i$.

\begin{figure}
\begin{center}
\includegraphics[width=0.9\linewidth]{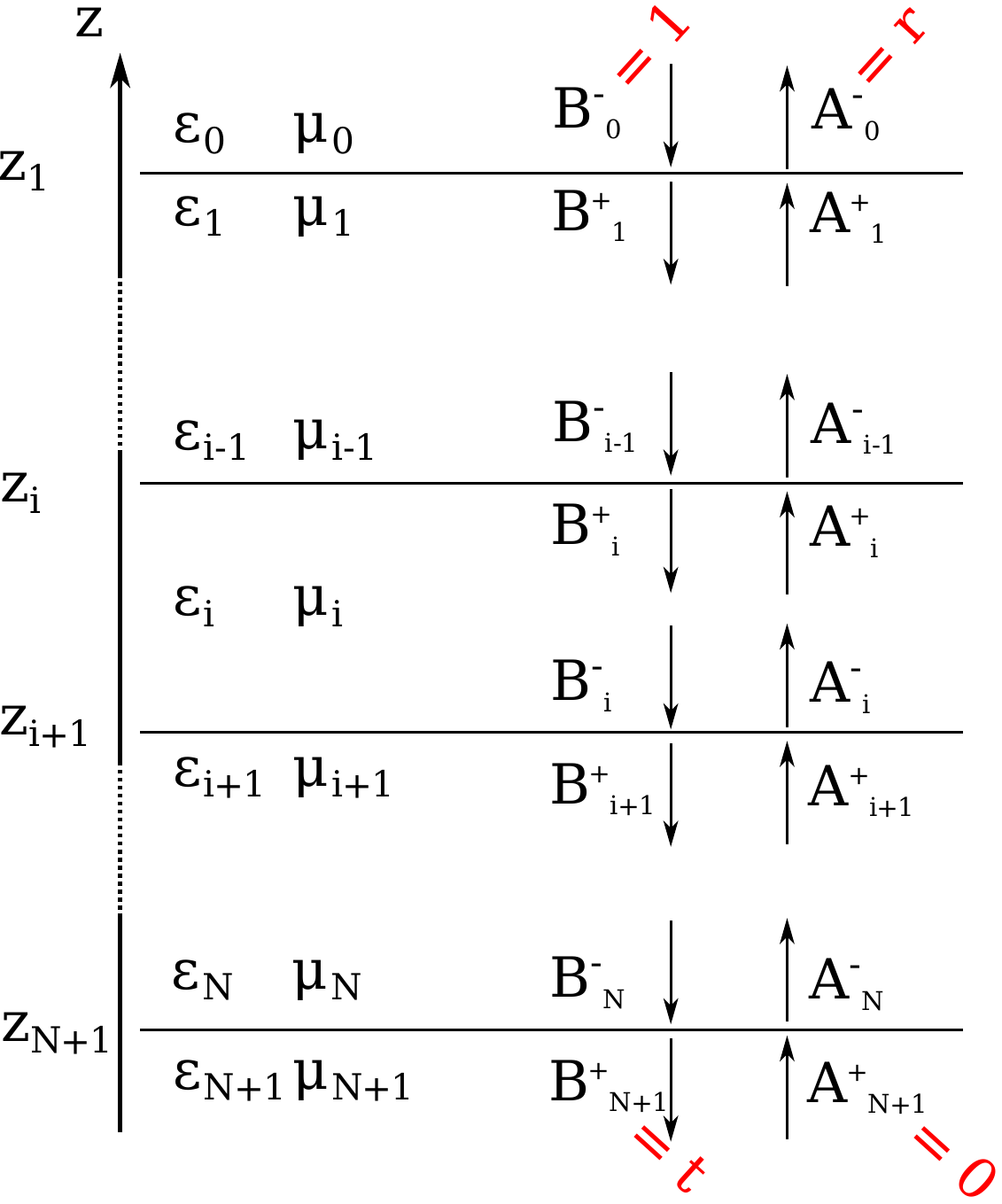}
\end{center}
\caption{Multilayer structure and definition of important variables. We consider the light coming from the top of the image. $\epsilon_i$ and $\mu_i$ are respectively the permittivity and the permeability of layer $i$. $A_i$ (resp. $B_i$) are defined as the amplitude of the upward propagating (resp. downward propagating) field in layer $i$, and the $+$ (resp. $-$) superscript indicates that the field value is taken at the top (resp. bottom) of layer $i$.}
\label{fig:multicouches}
\end{figure}

The solution to Eq.~\eqref{eq:Helmholtz} can be written in layer $i$ as:
\begin{equation}
    \begin{cases}
    E_{y,i}(z) &= A_i^+ e^{i \gamma_i (z-z_i)} + B_i^+ e^{-i \gamma_i (z-z_i) }\,,\\
    E_{y,i}(z) &=  A_i^- e^{i \gamma_i (z-z_{i+1})} + B_i^- e^{-i \gamma_i (z-z_{i+1}) }\,,        
    \end{cases}\label{form} 
\end{equation}

where $A_i^\pm$ (resp. $B_i^\pm$) are defined as the amplitude of the upward propagating (resp. downward propagating) field in layer $i$, and the $+$ (resp. $-$) superscript indicates that the phase reference of each mode is taken at the top (resp. bottom) of layer $i$, as the equations show.
Figure~\ref{fig:multicouches} represents these coefficients in the corresponding layers and close to their respective interfaces.

The continuity conditions for the tangential electric and magnetic fields at each interface between layer $i$ and layer $i+1$ lead to the following equations:
\begin{align}
    A_{i}^- + B_i^- &= A_{i+1}^+ + B_{i+1}^+\, , \label{eq:cont1}\\
    \frac{\gamma_i}{\mu_i}(A_{i}^- - B_i^-) &= \frac{\gamma_{i+1}}{\mu_{i+1}} (A_{i+1}^+ - B_{i+1}^+) \label{eq:cont2}\,.
\end{align}

When the amplitude of the incident plane wave is taken equal to 1 ($B_0^- = 1$) and there is no incoming light from below ($A_{N+1}^- = 0$), then the reflection and transmission coefficient are $A_0^- = r$ and $B_{N+1}^+ = t$, respectively. This means we have a system of $2(N+1)$ continuity equations and $2N$ propagation equations for just as many unknowns (the four $A_i^\pm$ and $B_i^\pm$ coefficients of each layer $i\in[1;N]$, $A_0^-$ and $B_{N+1}^+$). 
In theory, such a system could be solved numerically using classical algorithms.
However, the corresponding matrix is sparse and contains figures with very different magnitudes so that classical approaches are numerically unstable, to the point that it often can not be used for more than typically a few layers. Specific formalisms to solve this system of equations have therefore been developed, and are presented below.

The TM polarization can be treated simply by changing $\mu_i$ into $\epsilon_i$ in the system above. We underline however that there is a $\pi$ phase in normal incidence for the reflection coefficient defined in TE and in TM polarization.

Now, because of Eq.~\ref{eq:gamma}, we need to discuss how the square root of a complex number is defined and how this may have an impact on the accuracy of the calculation. By default, on any computer, the phase of a complex number $\zeta$ is defined in $\left[-\pi,\pi\right[$, meaning that taking its square root thus produces a complex number with an argument in $\left[-\pi/2,\pi/2\right[$, \textit{i.e.,} {\em with positive real part and arbitrary imaginary part}.
This determination of the square root can be changed manually, in order to obtain a positive imaginary part and arbitrary real part for instance, by taking the opposite of the result of the square root if the argument of $\zeta$ is negative.
In any case, this introduces a cut in the complex plane, which cannot always be ignored.

Here, Eq.~\eqref{form} indicates that changing the determination of the square root for $\gamma$ is equivalent to exchanging $A$ and $B$ {\em for all the layers in the structure}, so that this is not expected to change the result of the computation, as Eq.~\eqref{eq:cont1} and Eq.~\eqref{eq:cont2} are unchanged by this permutation. However, it is clear that if $\gamma_i$ has a negative imaginary part, then these equations introduce positive real exponentials, which would amplify any numerical error made on their amplitude.
We have made the choice to use $\gamma_i$ with positive imaginary parts in PyMoosh to maximize the stability of the code by using only decreasing exponential functions in all our calculations (especially the computation of the whole field, see below), inspired by what is done in RCWA \cite{lalanne_highly_1996, granet_efficient_1996}.
The change we have brought to the determination of the square root for the surrounding medium has more importance and will be discussed later.

From there, there are different ways to eliminate all the unknowns to obtain the reflection or the transmission coefficients. The most common formalism is based on Transfer Matrices (T-matrices). However, a more physical and stable formalism consists in writing reflection and transmission coefficients for each layer or interface. This is the Scattering Matrix  (S-Matrix) method, widely used in more advanced simulation techniques where evanescent waves are ubiquitous. Since the work of Abélès\cite{Abeles1950theorie}, we also know that considering the impedance or admittance of each layer is a relevant approach. This approach can be translated both into a matrix formalism and into a recursive formula to compute how the admittance of a structure is modified by adding a supplementary layer\cite{macleod2017thin}. Because these methods are not perfectly stable, an adaptation called the Dirichlet-to-Neumann (DtN) maps has been proposed more recently\cite{hughes1995multiscale}. The DtN technique is related to the Abélès matrices formalism exactly in the same way that the scattering matrix formalism is related to the transfer matrix approach.
In the following we derive the main formalisms of these different approaches.

\subsection{T-matrices}

Let us define $\psi_i = \frac{\gamma_i}{\mu_i}$ for all layers.
For the interface between layers $i$ and $i+1$, the continuity conditions \eqref{eq:cont1} and \eqref{eq:cont2} can be rewritten as:
\begin{align*}
    \begin{pmatrix}
        A_{i+1}^+ \\
        B_{i+1}^+
    \end{pmatrix}
     &= T_{i, i+1}
    \begin{pmatrix}
        A_{i}^- \\ B_{i}^-
    \end{pmatrix}\\
    &= \frac{1}{2 \psi_{i+1}}
    \begin{pmatrix}
        \psi_{i+1} + \psi_{i} & \psi_{i+1} - \psi_{i}\\ 
        \psi_{i+1} - \psi_{i} & \psi_{i+1} + \psi_{i}
    \end{pmatrix}
    \begin{pmatrix}
        A_{i}^- \\ B_{i}^-
    \end{pmatrix},
\end{align*}
which defines the T-matrix for a given interface. The propagation relations
linking $A_i^-, B_i^-$ and $A_i^+, B_i^+$, through the layer of height $h_i$ can be written as 
\begin{equation}
    \begin{pmatrix}
    A_{i}^- \\ B_{i}^-
    \end{pmatrix}
     = C^{(T)}_{i}
    \begin{pmatrix}
        A_{i}^+ \\ B_{i}^+
    \end{pmatrix}
    =
    \begin{pmatrix}
        e^{-i\gamma_{i}h_i}  & 0\\ 
        0 & e^{i\gamma_{i}h_i}
    \end{pmatrix}
    \begin{pmatrix}
        A_{i}^+ \\ B_{i}^+
    \end{pmatrix}\,,
\end{equation}
which defines the T-matrix for layer $i$. We underline that two exponential functions appear in the above expression. When the propagation constant $\gamma_i$ is complex, one of these two terms can be very small in front of the other, leading to floating point cancellation\cite{muller2018handbook} and to numerical errors.
Layer and interface matrices just need to be multiplied by each other to eliminate the unknowns corresponding to the field amplitude inside each layer, but this is what may produce floating point cancellation:

\begin{equation}
    \begin{pmatrix}
    A_{N+1}^+ \\ B_{N+1}^+
    \end{pmatrix}
     =  T_{N, N+1} \prod_{i=0}^{N-1} \left( C^{(T)}_{i+1} T_{i, i+1} \right)
    \begin{pmatrix}
        A_{0}^- \\ B_{0}^-
    \end{pmatrix}\,.
\end{equation}

From the transfer matrix 
\begin{equation}
    T = T_{N, N+1} \prod_{i=0}^{N-1} \left( C^{(T)}_{i+1} T_{i, i+1} \right)\,,
\end{equation}
for the whole structure we can retrieve coefficients \begin{equation}
r = A_0^- = -\frac{T_{01}}{T_{00}}\end{equation} and \begin{equation}t = B_{N+1}^+ = r T_{10} + T_{11}.\end{equation}

\subsection{S-matrices}

Scattering matrices are simply another way to eliminate the unknowns. The S-matrix formalism uses the same field definitions, but rewrites the continuity conditions obtained from Eqs.~\eqref{eq:cont1}-\eqref{eq:cont2} in order to link incoming fields at the interface $(A_i^+, B_{i+1}^-)$ with the outgoing fields $(A_{i+1}^-, B_{i}^+)$, giving the new interface matrices:
\begin{align}
    \begin{pmatrix}
        A_{i}^- \\
        B_{i+1}^+
    \end{pmatrix}
     &= I_{i, i+1}
    \begin{pmatrix}
        B_{i}^- \\ A_{i+1}^+
    \end{pmatrix}\\
    &= \frac{1}{\psi_{i} + \psi_{i+1}}
    \begin{pmatrix}
        \psi_{i} - \psi_{i+1} & 2 \psi_{i+1}\\ 
        2 \psi_{i}  & \psi_{i+1} - \psi_{i}
    \end{pmatrix}
    \begin{pmatrix}
        B_{i}^- \\ A_{i+1}^+
    \end{pmatrix}.\label{eq:Smat_S}
\end{align}

The propagation matrices are also modified:
\begin{equation}
    \begin{pmatrix}
    A_{i}^+ \\ B_{i}^-
    \end{pmatrix}
     = C^{(S)}_{i}
    \begin{pmatrix}
        B_{i}^+ \\  A_{i}^- 
    \end{pmatrix}
    =
    \begin{pmatrix}
        0 & e^{i\gamma_{i}h_i}\\ 
        e^{i\gamma_{i}h_i} & 0
    \end{pmatrix}
    \begin{pmatrix}
        B_{i}^+ \\  A_{i}^-
    \end{pmatrix}\,.\label{eq:Smat_C}
\end{equation}

As we will see later, these propagation matrices, where both exponentials are identical, lead to a greater numerical stability. However, computing the matrix linking boundary conditions is no longer a straightforward matrix product, since the coefficients of a given layer are not on the same side of Eq.~\eqref{eq:Smat_S}. Therefore, the relationship between the different $I$ and $C^{(S)}$ matrix equations can be written as:

\begin{equation}
    \begin{pmatrix}
    A \\ B
    \end{pmatrix}
    =
    \begin{pmatrix}
        U_{00} & U_{01}\\ 
        U_{10} & U_{11}
    \end{pmatrix}
    \begin{pmatrix}
        C \\  D
    \end{pmatrix} \text{, and }
    \begin{pmatrix}
    D \\ E
    \end{pmatrix}
    =
    \begin{pmatrix}
        V_{00} & V_{01}\\ 
        V_{10} & V_{11}
    \end{pmatrix}
    \begin{pmatrix}
        B \\  F
    \end{pmatrix}\,,
\end{equation}
where the new matrix we want to define is:
\begin{equation}
    \begin{pmatrix}
    A \\ E
    \end{pmatrix}
    =
    \begin{pmatrix}
        S_{00} & S_{01}\\ 
        S_{10} & S_{11}
    \end{pmatrix}
    \begin{pmatrix}
        C \\  F
    \end{pmatrix}\,.
\end{equation}

We can define a "cascading" operation, computing $S$ as a function of $U$ and $V$ coefficients:
\begin{equation}
    S
    =
    \begin{pmatrix}
        U_{00} + \frac{U_{01}V_{00}U_{10}}{1-V_{00}U_{11}} & \frac{U_{01}V_{01}}{1-V_{00}U_{11}}\\ 
        \frac{U_{10}V_{10}}{1-V_{00}U_{11}} & V_{11} + \frac{V_{10}U_{11}V_{01}}{1-V_{00}U_{11}}
    \end{pmatrix}\,.
    \label{eq:cascade_Smat}
\end{equation}

Applying this computation iteratively on every interface and layer leads to the scattering matrix $S$ for the whole structure:
\begin{equation}
    \begin{pmatrix}
    A_{0}^- \\ B_{N+1}^+
    \end{pmatrix}
     = S
    \begin{pmatrix}
        B_{0}^- \\  A_{N+1}^+
    \end{pmatrix}\,.\label{eq:Smat_tot}
\end{equation}

The reflection and transmission coefficient can be read directly, knowing $ B_{0}^-=1\,,  A_{N+1}^+=0$: $r=S_{00}\,, t=S_{10}$.

\subsection{Abélès formalism}

Contrary to T-matrices and S-matrices, the Abélès formalism (as well as the following Dirichlet-to-Neumann formalism) uses only one $E_y$ field position in each layer, thus removing the need for propagation matrices. We will also therefore directly speak in terms of $E_{y,i}$ and $\partial z E_{y,i}$. Taking for instance the field on the upper side of the layers and using the field continuity, we can write:

\begin{equation}
\begin{cases}
    E_{y,i}(z_{i}) &= A_i^+ + B_i^+\,, \\
    \partial z E_{y,i}(z_{i})/\mu_i &= i\psi_i (A_i^+ - B_i^+) \,,\\
    E_{y,i+1}(z_{i+1}) &= A_i^+ e^{-i\gamma_i h_i} + B_i^+ e^{i\gamma_i h_i} \,,\\
    \partial z E_{y,i+1}(z_{i+1})/\mu_{i+1} &= i\psi_i (A_i^+ e^{-i\gamma_i h_i} - B_i^+ e^{i\gamma_i h_i})\,.
\end{cases}
\end{equation}

These relations and the continuity conditions at the interface at $z_{i+1}$ lead to the matrix equations:
\begin{align*}
    \begin{pmatrix}
    E_{y,i+1} \\ \partial z E_{y,i+1}/\mu_{i+1}
    \end{pmatrix}
    & = M_{i}
    \begin{pmatrix}
    E_{y,i} \\ \partial z E_{y,i}/\mu_{i}
    \end{pmatrix}\\
    &=
    \begin{pmatrix}
        \cos(\gamma_i h_i)  & -\frac{\sin(\gamma_i h_i)}{\psi_i}\\ 
        \psi_i\sin(\gamma_i h_i) & \cos(\gamma_i h_i)
    \end{pmatrix}
    \begin{pmatrix}
    E_{y,i} \\ \partial z E_{y,i}/\mu_{i}
    \end{pmatrix}.
    \label{eq:Abélès}
\end{align*}

 Once the matrix product $M = \prod_i M_i$ is computed, the $r$ and $t$ coefficients are slightly more complicated to retrieve than for T-matrices: 
\begin{align}
    E_{y,N+1} &= t \\
    E_{y,0} &= r+1 \\
    \partial z E_{y,N+1}/\mu_{N+1} &= -i\psi_{N+1} t\\
    \partial_z E_{y,0}/\mu_{i=0} &= i\psi_0 (r-1)
\end{align}

\subsection{Dirichlet-to-Neumann formalism}

Finally, just like the S-matrix formalism was the equivalent of T-matrices but using incoming field on one side of the equation and outgoing fields on the other, the Dirichlet-to-Neumann (DtN) formalism is the equivalent of the Abélès formalism but keeping $E_y$ on one side of the equation and $\partial z E_y$ on the other:
\begin{align}
    \begin{pmatrix}
    \partial zE_{y,i}/\mu_i \\ \partial_z E_{y,i+1}/\mu_{i+1}
    \end{pmatrix}
     &=
    N_{i, i+1}
    \begin{pmatrix}
    E_{y,i} \\  E_{y,i+1}
    \end{pmatrix}\\
    &=
    \begin{pmatrix}
        \frac{\psi_i \cos(\gamma_i h_i)}{\sin(\gamma_i h_i)}  & -\frac{\psi_i}{\sin(\gamma_i h_i)}\\ 
        \frac{\psi_i}{\sin(\gamma_i h_i)} &  -\frac{\psi_i \cos(\gamma_i h_i)}{\sin(\gamma_i h_i)}
    \end{pmatrix}
    \begin{pmatrix}
    E_{y,i} \\  E_{y,i+1}
    \end{pmatrix}.\label{eq:D2N}
\end{align}

As with S-matrices, the complete structure matrix computation is not straightforward, and needs a cascading operation similar to the one used for S-matrices. To this end, if we call $A = N_{i, i+1}$ and $B=N_{i+1, i+2}$, then the $D_{i, i+2}$ matrix linking fields in layer $i$ and in layer $i+2$ is given by:

\begin{equation}
    D_{i, i+2} = 
    \begin{pmatrix}
         A_{0 0} - \frac{A_{0 1}  A_{1 0}}{A_{11} - B_{00}} &  \frac{A_{0 1}  B_{0 1}}{A_{11} - B_{00}}\\
         - \frac{B_{1 0}  A_{1 0}}{A_{11} - B_{00}} & B_{1 1} + \frac{B_{1 0}  B_{0 1}}{A_{11} - B_{00}}
    \end{pmatrix}\,.
\end{equation}

For all other formalisms, we typically consider a superstrate of thickness 0, in order to have a phase reference exactly at the interface between the superstrate and the first layer of the structure. However, for the DtN formalism, as we can see in Eq.~\eqref{eq:D2N} this would lead to division by 0 in the first matrix. So we choose an arbitrary thickness for the superstrate ($\lambda/100$ in the PyMoosh code), and compensate for the additional phase at the end of the computation.

\subsection{Admittance formalism}

The admittance formalism is based on the Abélès formalism and its main advantage is to make  computations of the structure's reflectivity much faster.
We define:
\begin{equation}
    \delta_i = \gamma_i h_i, \,\,\,\text{and}\,\,\, n_{eff,i} = \frac{\gamma_i}{\mu_i k_0},
\end{equation}
with $\gamma_i$ as defined in Eq.~\eqref{eq:gamma}.

With the Abélès formalism (inverting Eq.~\eqref{eq:Abélès}):
\begin{equation}
    \begin{pmatrix}
    E_{y,i} \\  \partial_z E_{y,i}/\mu_i
    \end{pmatrix}
     =
    \begin{pmatrix}
\cos \delta_i & \frac{\sin \delta_i}{\psi_{i}} \\
-\psi_i \sin \delta_i & \cos \delta_i
    \end{pmatrix}
    \begin{pmatrix}
    E_{y,i+1} \\ \partial_Z E_{y,i+1}/\mu_{i+1}
    \end{pmatrix}\,.
\end{equation}

We then define $X_i = \frac{\partial_z E_{y,i}}{\mu_{i}E_{y,i}}$:
\begin{equation}
    \begin{pmatrix}
    \frac{E_{y,i}}{E_{y,i+1}} \\  \frac{\partial_z E_{y,i}}{\mu_{i}E_{y,i+1}}
    \end{pmatrix}
     =
    \begin{pmatrix}
\cos \delta_i & \frac{\sin \delta_i}{\psi_{i}} \\
-\psi_i \sin \delta_i & \cos \delta_i
    \end{pmatrix}
    \begin{pmatrix}
    1 \\ X_{i+1}
    \end{pmatrix}\,.
\end{equation}

Computing the matrix-vector product and dividing both rows, we get:
\begin{equation}
    X_i = \frac{X_{i+1} - \psi_i \tan \delta_i}{1 + \frac{X_{i+1}}{\psi_i} \tan \delta_i}\,.
\end{equation}

Let us define $Y_i = \frac{i}{k_0}X_i$, leading to:

\begin{equation}
    Y_i = \frac{Y_{i+1} - i n_{eff,i} \tan \delta_i}{1 - i \frac{Y_{i+1}}{n_{eff,i}} \tan \delta_i}\,.
\end{equation}

Then, we use $Y_N = n_{eff, N}$ as initial condition and compute all $Y_i$, with the final reflection coefficient being for TE polarization:
\begin{equation}
r_{\text{TE}} = \frac{n_{eff,0} - Y_0}{n_{eff,0} + Y_0}\,.    
\end{equation}

In TM, the difference is that $n_{eff,i}^{(TM)} = \frac{\epsilon_i k_0}{\gamma_i}$ and 

\begin{equation}
r_{\text{TM}} = -\frac{ n_{eff,0}^{(TM)} - Y_0}{ n_{eff,0}^{(TM)} + Y_0}\,.    
\end{equation}

This formalism allows to compute the reflectance of a structure very quickly. It can also be generalized to transmission computations, but then loses its computation speed advantage. It can be however of particular interest for pedagogical reasons or in time-critical applications, given its simplicity and, as a direct consequence, its computational efficiency.

\subsection{Comparing the different methods}

Choosing the right method to compute the reflectance or transmittance of a structure depends on the context and whether stability or 
computational speed are particularly sought after. No method is both the fastest and the most stable. We have thoroughly tested the stability and estimated the efficiency of each of the methods as implemented in PyMoosh, to assist users in determining the most suitable approach for their specific needs.

\begin{figure}[ht]
    \centering
    \includegraphics[width=0.8\linewidth]{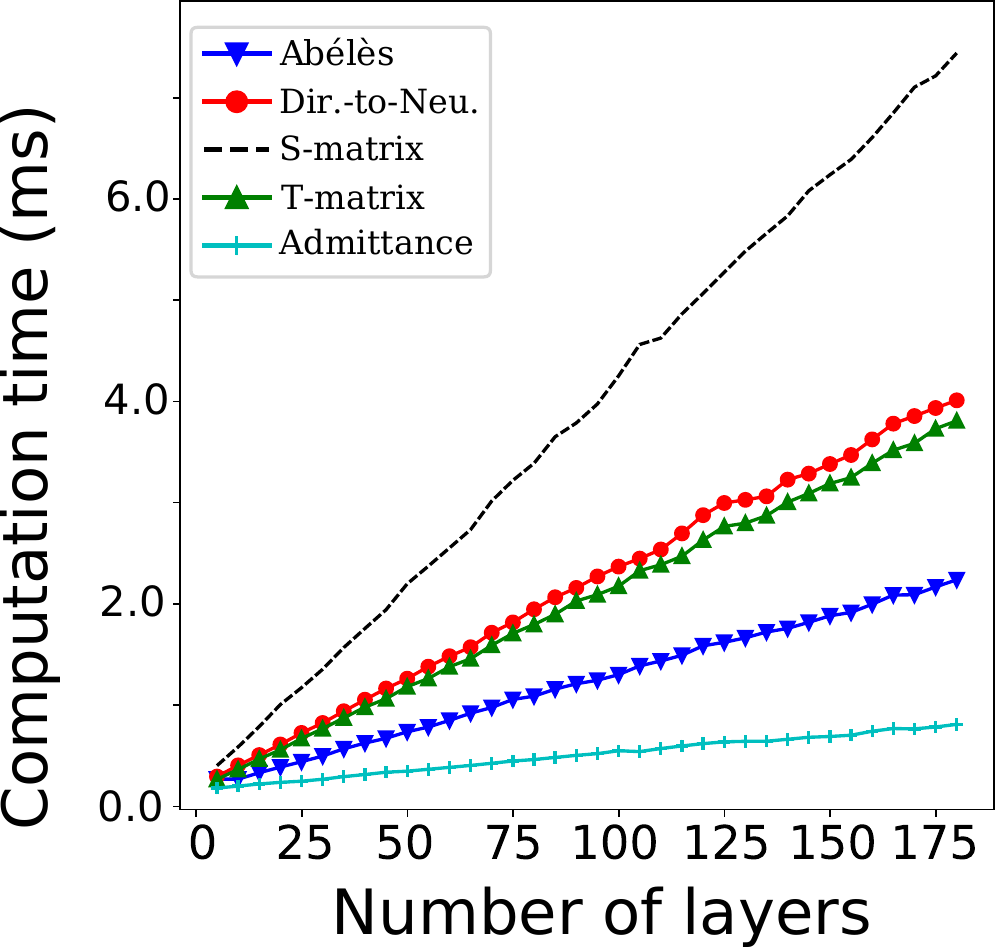}
    \caption{Computation time of the $r$ and $t$ coefficients for all methods as a function of the number of layers in a stack, averaged over 500 runs. As exact computation times are heavily influenced by hardware capabilities, it remains important to prioritize the relative comparison between formalisms over these precise times.}
    \label{fig:speed_bench}
\end{figure}

{\bf Computational speed}. As can be guessed from the mathematical expressions above and as is shown Fig.~\ref{fig:speed_bench}, the computational time required for computing the reflectance of a given structure increases linearly with the number of considered layers.
Also, as can be expected, the scattering matrix method is the slowest. It is typically twice as slow as the T-matrix or DtN methods.
On the other hand, the Abélès formalism is twice as fast as these, but the most computationally efficient technique of all is the admittance formula. It is thus ideally suited for optimizing the reflectance of a multilayer. Alas, this technique is also the most prone to stability issues. 

{\bf Numerical stability}. For a random dielectric structure, all formalisms can be considered stable, as the transmission can be expected to be relatively high. Numerical accuracy issues occur when evanescent waves are present, whether they are due to total internal reflection, to the presence of metals, or if evanescent Bloch waves are excited in the bandgaps of Bragg mirrors. While for optical filters such problems are unlikely, evanescent waves are ubiquitous in prism couplers or plasmonic structures, not to mention photonic crystal-based structures. 

\begin{figure}[ht]
    \centering
    \includegraphics[width=\linewidth]{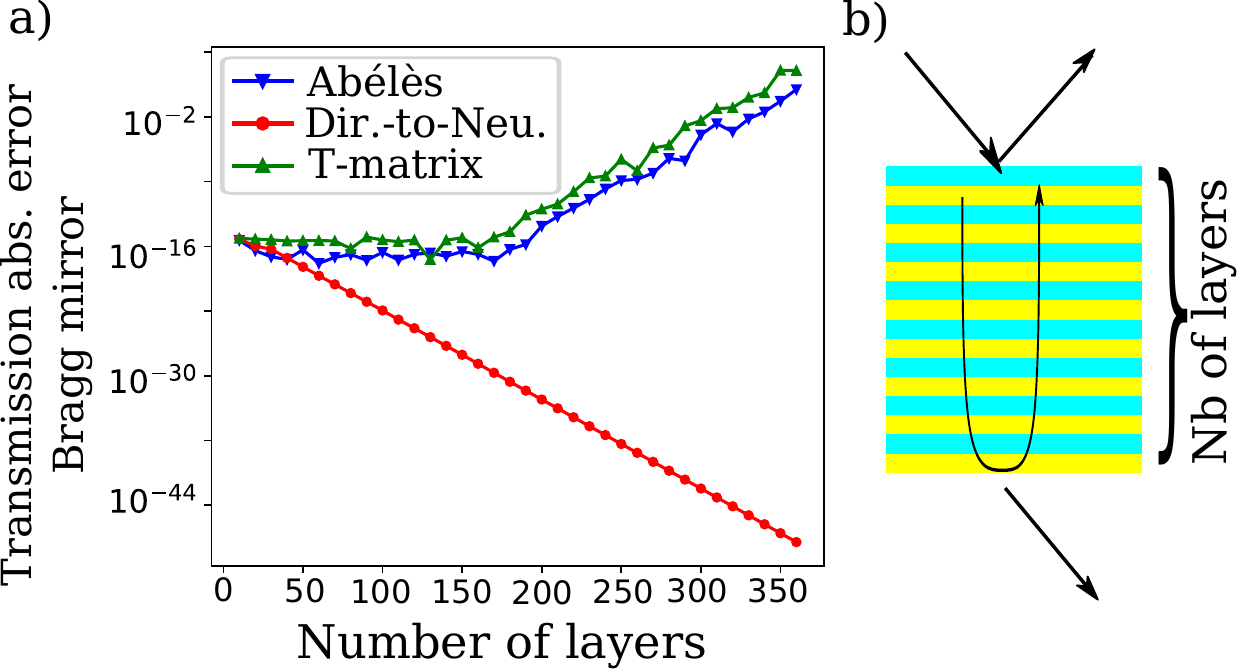}
    \caption{a) Absolute error between S-matrix (considered as our reference) and other formalisms.  Transmission coefficient for a quarterwave stack of materials $n=1.2$ and $n=1.5$ as a function of the total number of layers of the stack. b) Schematic representation of the structure.}
    \label{fig:stability_bench_Bragg}
\end{figure}

\begin{figure}[ht]
    \centering
    \includegraphics[width=\linewidth]{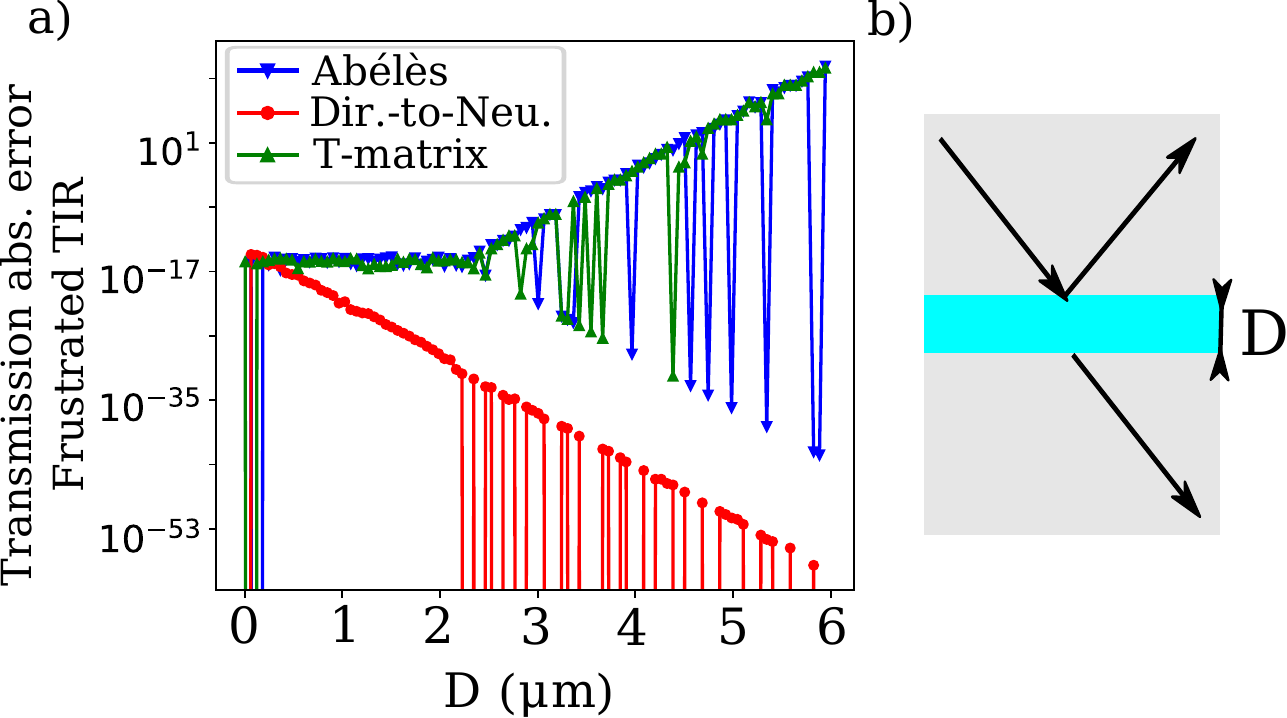}
    \caption{a) Absolute error between S-matrix (considered as our reference) and other formalisms. Transmission coefficient in a frustrated total internal reflection situation for a layer of air of width $D$ between materials of index $n=1.5$.  b) Schematic representation of the structure.}
    \label{fig:stability_bench_FTIR}
\end{figure}

\begin{figure}[ht]
    \centering
    \includegraphics[width=\linewidth]{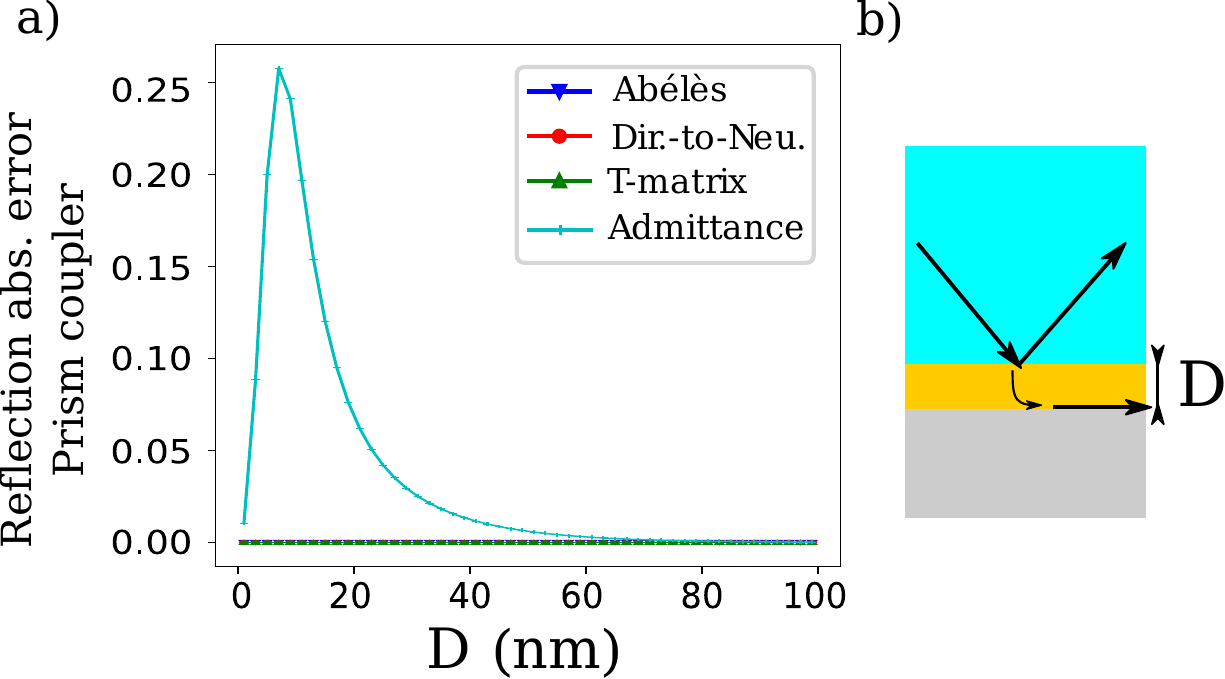}
    \caption{a) Absolute error between S-matrix (considered as our reference) and other formalisms. Surface plasmon resonance coupler in the Kretschmann-Raether configuration with a prism of index $n=1.5$ coupling to air through a silver layer of width $D$.  b) Schematic representation of the structure.}
    \label{fig:stability_bench_prism}
\end{figure}

The scattering matrix method has proven to be the most stable and accurate of all, regardless of the situation\cite{lalanne_highly_1996,granet_efficient_1996,bayPyLlamaStableVersatile2022}. For this reason, we consider its results as our reference in all the following. Figures~\ref{fig:stability_bench_Bragg}-\ref{fig:stability_bench_prism} show the absolute difference between the values given by the scattering matrix method and the other methods for the reflection and transmission coefficient in several cases. We discuss here explicitly the computation errors for TE illumination, but TM polarization faces roughly the same instabilities.

The first case, depicted in Fig.~\ref{fig:stability_bench_Bragg}, is a simple Bragg mirror, illuminated at oblique incidence ($15^\circ$) with a wavelength in the middle of the bandgap. As the number of layers increases, the transmission coefficient goes down to zero. As can be seen, the DtN and scattering matrix methods  differ only by an error of the order of machine precision. Therefore when the relative error stays the same, as the value of the transmission coefficient decreases, so does the absolute error.

It can be seen however that the Abélès formalism and the T-matrix method are subject to numerical instability, which typically occurs for more than 150 layers. This means that these methods should be considered with caution when structures involve Bragg mirrors, for instance in microcavities\cite{solnyshkov2021microcavity} for which the transmission through the stack is crucial. It should also be noted that the TE case is slightly more stable than the TM case, when no magnetic materials are involved, and that the Abélès formalism is significantly more stable under normal incidence illumination.

The second case, depicted in Fig.~\ref{fig:stability_bench_FTIR}, is a simple frustrated total internal reflection, \textit{i.e.,} a layer of air comprised between two layers of high index material $n=1.5$, illuminated with a $42^\circ$ incidence angle, slightly beyond the critical angle. When the air layer thickness is increased beyond a few wavelengths, the evanescent field within causes numerical instabilities for the T-matrices and the Abélès matrices, similar to what occurs for the Bragg mirror. This means these two algorithms should be avoided when simulating prism couplers, where evanescent waves occur. Above a thickness of 2 µm for the air layer, as the theoretical transmission tends towards zero, the DtN and S matrix methods often agree down to machine precision, leading to a zero absolute numerical error. They differ only by a relative error of the order of machine precision in other cases.

As a final example, we consider a metallic (silver) layer comprised between an upper layer with a high index ($n=1.5$) and air -- a typical prism coupler for a surface plasmon in Kretschmann-Raether configuration\cite{kretschmannNotizenRadiativeDecay1968}.
In the previous cases, we underline that the reflection 
coefficient is computed with no instability for T, S, Abélès or DtN matrices, even when the transmission coefficient is unstable. However, as can be seen in Fig. \ref{fig:stability_bench_prism}, 
the admittance formula is not accurate for thin metallic layers, and while the instability is not exponential, the value computed is not at all accurate. This shows that the admittance formalism is most adapted to optical filters with no metallic layers. While this can not be seen in the linear scale graph, shown in Fig. \ref{fig:stability_bench_prism}a, the difference between the values computed using the three other methods is down to machine precision.

Despite their stability in these typical examples, S and DtN matrices also have specific breaking points. For scattering matrices, when $\gamma_i/\mu_i \sim -\gamma_{i+1}/\mu_{i+1}$ (in TE polarization, but this occurs in TM too with $\epsilon_i$ instead of $\mu_i$), a division by zero occurs. This can only happen when lossless metamaterials are considered, which is not realistic, but is often studied for educational purposes (showing the ideal response of Pendry's lens for instance\cite{pendry_negative_2000}).
As for the DtN formalism, its better stability is due to the terms in $1/\sin$ and $1/\tan$, which are less prone to numerical errors than exponentials. However, there are still specific breaking points. For instance, for a layer of optical thickness $\lambda/2n$, where $n$ is the refractive index of the layer, in normal incidence, the argument of the $1/\sin$ is $\pi$, leading to a division by 0. As this may occur with any kind of transparent material, and it is actually typically used as a cavity in Distributed-Feedback Lasers, for instance, the DtN formalism must be used with care.

{\bf Discussion}. From the above results, it seems clear that scattering matrices should be the default choice when stability matters because it presents only very specific breaking points and we underline that the supplementary computational effort is reasonable. Transfer matrices seem an obvious choice in the literature, however our results show that the Abélès formalism is equivalently stable but twice as computationally efficient, while being no more complex to implement.
When speed really matters, when no metal is present and when the computation of reflectance is sufficient, the admittance formalism is a less explored but very interesting alternative, for instance in the context of optimization\cite{barry2020evolutionary} or for database generation for deep learning, being one order of magnitude more efficient than S-matrices. Finally, the Dirichlet-to-Neumann formalism is a good compromise for large multilayer stacks that may exhibit evanescent fields, as long as no $\lambda/2n$ layer is present. PyMoosh allows to make such choices and to assess easily the accuracy of any method by comparing it to the S-Matrix result.

\section{Computing absorption in an arbitrary layer}

In many applications, \textit{e.g.,} sensing or photovoltaic stacks, knowing where the absorption takes place is of paramount importance. This requires to compute the field amplitudes $A_i^\pm$ and $B_i^\pm$ that are ignored when only reflectance is to be computed. This can be done in any of the four main formalisms, and the present section describes how these fields can be computed for the S-matrix and the Abélès formalisms, because they have been shown to be respectively the most stable and the fastest. The admittance formalism can obviously not be used as it is designed to only compute the external coefficients.

In order to compute how much of the incident power is absorbed in each layer, the PyMoosh library computes the (time-averaged) projection along the $z$ axis of the normalized Poynting vector at the top of every layer $i$, $N_i$. This projected Poynting vector is proportional to the power flux through the top interface of the layer.
Since we normalize all power by the incident power, then the difference in power given by the difference in Poynting vector $\mathcal{A}_i = N_i$ - $N_{i+1}$ is the fraction of the incident power absorbed in layer $i$.

\vspace*{0.5cm}

\subsection{Abélès matrices}

With $M_i$ defined in Eq.~\eqref{eq:Abélès}, we can define $M_{tot}^{(k)} = \prod_{i=1}^{k} M_{i-1}$ which allows us to compute the fields in the structure:

\begin{align}
    \begin{pmatrix}
        E_{y,k}\\
        \frac{\partial_z E_{y,k}}{\mu_{k}}
    \end{pmatrix} = M_{tot}^{(k)}
    \begin{pmatrix}
        E_{y,0} \\
        \frac{\partial_z E_{y,0}}{\mu_{k=0}}
    \end{pmatrix} = M_{tot}^{(k)}
    \begin{pmatrix}
        r+1 \\
        i(r-1)\psi_0
    \end{pmatrix}\,.
\end{align}

Knowing this, the equation used to compute the normalized Poynting vector at the top of one layer is:

\begin{equation}
    N_i = \mathbf{Re}\left( -i E_{y, i} \partial_z E_{y,i}^* \frac{1}{\psi_0 \mu_i^*} \right)\,,
\end{equation}
in the TE case, and where $\mu$ and $\epsilon$ are switched to get the TM case.

\subsection{S-matrices}

Using the cascading operation defined in Eq.~\eqref{eq:cascade_Smat}, we can introduce the $\odot$ operator, where $A \odot B$ is the result of the cascading operation. This allows to write more simply the partial scattering matrices $U^{(k)}$ and $D^{(k)}$ corresponding to scattering matrices of incomplete structures: 
\begin{equation}
\begin{cases}
    U^{(2i)} &= C^{(S)}_0 \odot I_{01} \odot C^{(S)}_1 ... \odot I_{i-1, i}\,,\\
    U^{(2i+1)} &= C^{(S)}_0 \odot I_{01} \odot C^{(S)}_1 ... \odot C^{(S)}_{i}\,,\\
    D^{(2i)} &= C^{(S)}_{i} \odot I_{i, i+1} ... \odot C^{(S)}_{N+1}\,,\\
    D^{(2i+1)} &=  I_{i, i+1} \odot C^{(S)}_{i+1} ... \odot C^{(S)}_{N+1}\,.
\end{cases}
\end{equation}
In the case when we consider the upper coefficients of layer $i$, this allows to write 
\begin{align}
    \begin{pmatrix}
        A_0^-\\
        B_i^+
    \end{pmatrix} &= U^{(2i)}
    \begin{pmatrix}
        B_0^- \\
        A_i^+
    \end{pmatrix}\,,\\
    \begin{pmatrix}
        A_i^+\\
        B_{N+1}^+
    \end{pmatrix} &= D^{(2i)}
    \begin{pmatrix}
        B_i^+ \\
        A_{N+1}^+
    \end{pmatrix}\,.
\end{align}

The same can be written for the lower coefficients using odd index $2i+1$. The idea here is that if we assume an incoming plane wave from above with a unity amplitude (\textit{i.e.,} $B_0^- = 1$ and $A_{N+1}^+=0$), we can use the intermediate matrix that usually appear when calculating the cascading formula to retrieve the unknowns:

\begin{widetext}
    
\begin{align}
    \begin{pmatrix}
        A_i^{+}\\
        B_i^{+}
    \end{pmatrix} = \frac{1}{1 - U^{(2i)}_{11}  D^{(2i)}_{00}}
    \begin{pmatrix}
         D^{(2i)}_{01} & U^{(2i)}_{10}  D^{(2i)}_{00}\\
         U^{(2i)}_{11}  D^{(2i)}_{01} & U^{(2i)}_{10}
    \end{pmatrix}
    \begin{pmatrix}
        A_{N+1}^+\\
        B_0^-
    \end{pmatrix}.\label{eq:S_mat_side}
\end{align}
\end{widetext}

Again, using odd index $2i+1$ instead of $2i$ allows to retrieve the $A_i^-$ and $B_i^-$ coefficients. 

The equation used to compute the normalized Poynting vector at a given interface, using the corresponding coefficients, is:
\begin{align}
    N_i &= -\mathbf{Re}\left( \frac{\psi_i^*}{\psi_0} (A_i^+ - B_i^+)^*\,(A_i^+ + B_i^+) \right)\,,
\end{align}
where $\gamma$ is defined in Eq.~\eqref{eq:gamma}, for the TE case. For the TM case, simply replace $\epsilon$ with $\mu$ as usual.

\subsection{Application to solar cells}

Solar cells can most often be modeled as a multilayered structure. Once the absorption in the active part, $\mathcal{A}_{active}$, is computed for the whole spectrum, it can be converted into an estimate of the short-circuit current $j_{sc}$ delivered by the cell under standard illumination (an AM1.5 solar spectrum) using the following formula, assuming a unity quantum yield \cite{santbergen2010am1}:

\begin{equation}\label{eq:jsc}
j_{sc} = \int \mathcal{A}_{active}(\lambda) \frac{d I}{d\lambda}.
 \frac{\mbox{e}\lambda}{\mbox{h} \mbox{c}}\,\mbox{d}\lambda\,,
\end{equation}
where $\frac{d I}{d\lambda}$ is the spectral intensity for the AM1.5 spectrum per wavelength, $e$ the charge of the electron, $h$ Planck's constant and $c$ the speed of light. The efficiency of the device can then be computed as the ratio of the short-circuit current over the maximum achievable current for the considered spectral window.

\begin{figure*}[ht]
    \centering
    \includegraphics[width=\linewidth]{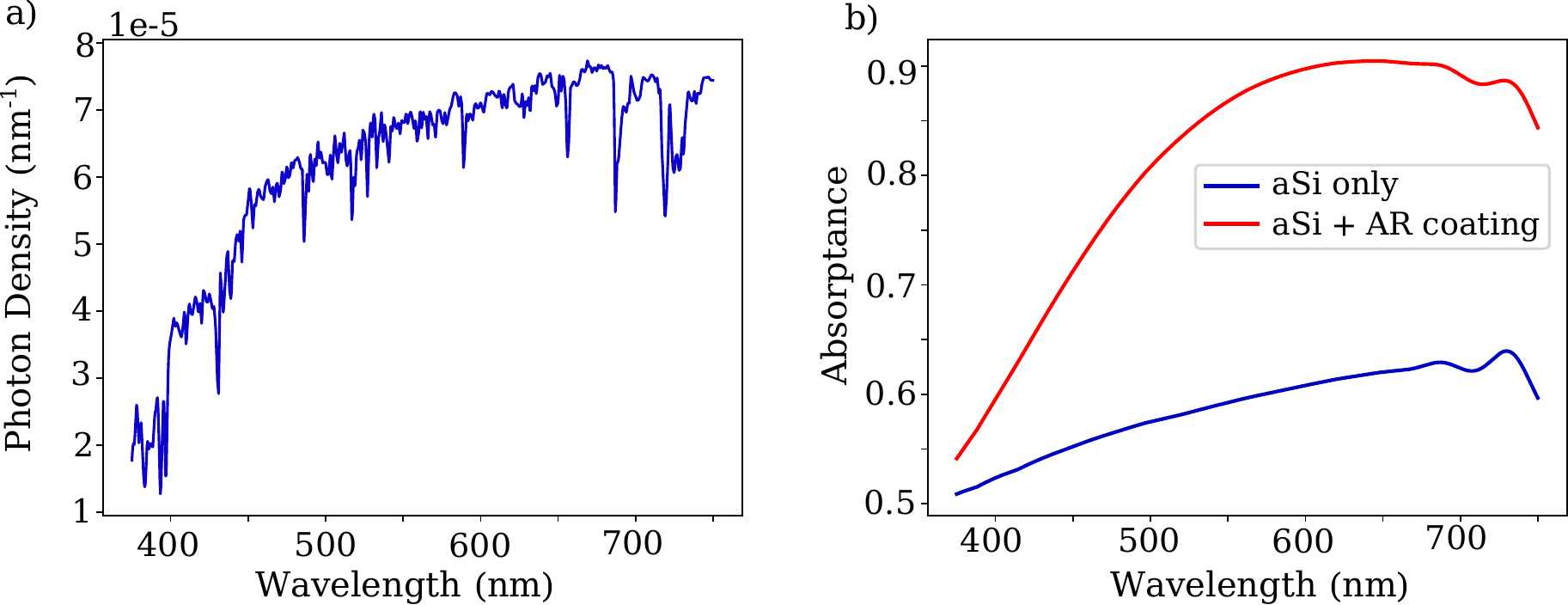}
    \caption{a) Photon spectral density (photons per wavelength unit) of the AM1.5 visible solar spectrum. b) Absorptance in the active layers of 2 different basic solar cells (S-Matrix). 1 $\upmu$m of SiA bulk with or without adding a layer of optical index $n=1.5$ and thickness $\lambda/4n$ with $\lambda = 600$~nm.}
    \label{fig:PV}
\end{figure*}

Because of how important this type of application is, there are functions in PyMoosh specifically designed to simulate photovoltaic multilayers, taking into account the photon density of the solar spectrum and the conversion efficiency of solar cells.

Figure~\ref{fig:PV}a) shows the photon density of the visible solar spectrum, received on the Earth, as included in the PyMoosh library. Figure~\ref{fig:PV}b) shows the absorptance of two structures: one that is a single layer of amorphous Silicon (aSi) of 1~$\upmu$m thickness, and a second structure with an additional dielectric anti-reflecting (AR) layer on top ($n=1.5, h=\lambda/4n, \lambda=600$~nm).
These absorptance curves show that bulk aSi already has an absorptance of about 50\%, and that a single AR coating can improve this absorptance to over 85\% for half of the visible spectrum. 

\section{Field computation}

While reflectance or transmittance spectra are among the most represented quantities, it is highly informative to be able to compute the electric or magnetic fields inside a structure in response to an illumination, either with an incoming beam, or when a source is placed in the structure. The magnetic field is especially relevant for all situations involving metals. The computational cost is less of a concern in such a situation, while the numerically stability is crucial. We thus use the S-matrix method in that context, and the technique explained above to compute all the intermediate coefficients $A_i^\pm$ and $B_i^\pm$. The electric (resp. magnetic) field in TE (resp. TM) polarization can be written, assuming $x=0$ and for a single incident plane wave:
\begin{equation}
    E_{y,i}(x=0,z) =  A_i^- e^{i \gamma_i (z-z_{i+1})} +B_i^+ e^{-i \gamma_i (z-z_i) }\,.\label{eq:fields}
\end{equation}

Using this expression and the determination of the square root mentioned above (positive imaginary part), presents the advantage that, since all the coefficients are computed with the best possible accuracy, all the exponentials are decreasing away from the interfaces. No numerical error can thus be amplified.

\subsection{Illumination with a Gaussian beam}

A Gaussian beam can be written as a superposition of plane waves. In TE polarization, the $E_y$ field in layer $i$ can be written as:
\begin{widetext}
\begin{equation}
E_{y,i} (x,z)  = \frac{1}{2\pi}\int \mathcal{E}_0(k_x) \left(A_i^- e^{i \gamma_i (z-z_{i+1})} +B_i^+ e^{-i \gamma_i (z-z_i)}\right)\,e^{ik_x x}\mbox{d} k_x, \label{eq:amplitude}
\end{equation}
where the amplitude is a Gaussian:
\begin{equation}
\mathcal{E}_0 \left(k_x \right) = \frac{w}{2 \sqrt{\pi}} e^{-\frac{w^2}{4}
\left(k_x-n k_0 \sin \theta_0 \right)^2} e^{- i k_x x_0},\label{eq:gaussian}
\end{equation}
\end{widetext}
for a Gaussian beam centered at $x_0$, $\theta_0$ being the central incidence angle and $w$ the waist of the beam. The refractive index of the upper medium is noted $n$. We underline that the above definition of $w$ is standard for the study of non-specular phenomena\cite{tamir1986nonspecular,polles2010light}, but that this means $w$ is the waist along the $x$ axis and not the minimum waist of the beam perpendicularly to the propagation direction ($w'$). There is a $\cos \theta$ factor between these two definitions: $w' = w/\cos\theta$. While for small angles of incidence this has little importance, for grazing angles it should be taken into account.

In Pymoosh, the field is computed for each given $k_x$ in all the layers using Eq.~\eqref{eq:fields} and for all the points defined in the vertical discretization.
Then, the full field at different horizontal positions is computed by multiplying this vertical vector by $e^{i k_x x}$ for all the values of $x$ included in the horizontal discretization. The thus computed fields for each plane wave $k_x$ are subsequently multiplied by the corresponding complex amplitude $\mathcal{E}_0 \left(k_x \right)$ (see Eq.~\eqref{eq:gaussian}) and all the plane wave contributions are summed,  leading to the calculation given by Eq.~\eqref{eq:amplitude}.
Importantly, as the number of incidence angles (and thus $k_x$ values) is finite and thus discretized, in consequence the resulting field is periodic. The period is what is called the window size, $d$. If $\Delta k_x$ is the difference between two successive values of $k_x$, we have the relation $d = \frac{2\pi}{\Delta k_x}$. The number of plane waves is chosen such that the amplitude for the maximum and minimum wavevectors given by Eq.~\eqref{eq:gaussian} is smaller than 1\% and has thus no impact on the final field-image. 

The exact same procedure can be applied to obtain images corresponding to $H_x$ and $H_z$ in TE polarization:
\begin{align}
H_x(0,z)  &= -\frac{\gamma_i}{\omega \mu_i}\left( A_i^- e^{i \gamma_i (z-z_{i+1})} - B_i^+ e^{-i \gamma_i (z-z_i) }\right)\,,\\
H_z(0,z)  &= \frac{k_x}{\omega \mu_i}\left( A_i^- e^{i \gamma_i (z-z_{i+1})} + B_i^+ e^{-i \gamma_i (z-z_i) }\right)\,,
\end{align}
and for the TM polarization, $H$ and $E$, $\mu$ and $\epsilon$ have to be exchanged. Theoretically a minus sign should also be added in front of all the expressions, adding a global phase-shift of $\pi$. Since the phase origin is arbitrary, we did not take it into account in the code.

\subsection{Current line source}

\begin{figure}
    \centering
    \includegraphics[width=0.5\linewidth]{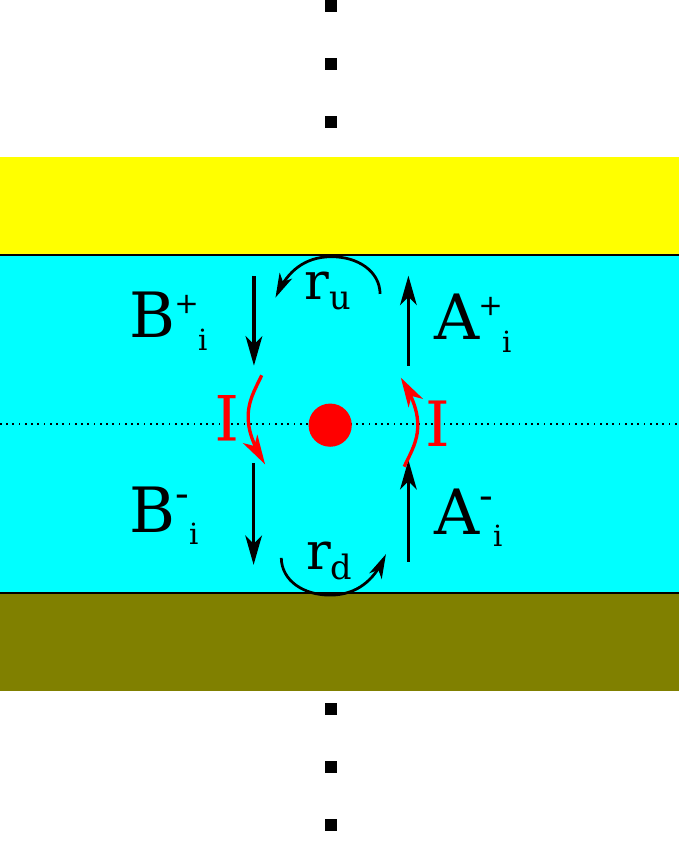}
    \caption{Schematic representation of the field values surrounding the source.}
    \label{fig:field_schematic}
\end{figure}

It is relatively simple to compute the field corresponding to a source of oscillating current $I$, invariant in the $y$ direction, in TE polarization (electric field, hence the oscillation direction, along the infinite axis), even in a complex multilayered structure. The equation that must be solved inside the homogeneous layer where the source is placed, can be written:
\begin{equation}
    \frac{\partial^2 E_y}{\partial z^2} 
    +    \frac{\partial^2 E_y}{\partial x^2} 
    +\epsilon_i\mu_i k_0^2 E_y = - i\omega \mu_0\mu_i I \delta(x-x_s,z)\,,
\end{equation}
with $\delta$ the Dirac function, if we assume that the source is placed in $(x_s,0)$. Because of the invariance of the problem, taking the Fourier transform in the $x$ direction yields:
\begin{equation}
    \frac{\partial^2 \tilde{E}_y}{\partial z^2} 
    +(\epsilon_i\mu_i k_0^2 - k_x^2) \tilde{E}_y = - i\omega \mu_0\mu_i I \delta(z) \, e^{i k_x x_s} \, .
\end{equation}
This means that, while $\tilde{E}_y$ is continuous, its derivative is not, but undergoes a jump equal to $ - i\omega \mu_0\mu_i I e^{i k_x x_s}$. We assume a single source in layer $i_s$, above which the field can be written:
\begin{equation}
    \tilde{E}_{y,i_s}(z) = A_{i_s}^+ e^{i \gamma_{i_s} z} +B_{i_s}^+ e^{-i \gamma_{i_s} z } \, ,
\end{equation}
and below it is:
\begin{equation}
    \tilde{E}_{y,i_s}(z) = A_{i_s}^- e^{i \gamma_{i_s} z} +B_{i_s}^- e^{-i \gamma_{i_s} z } \, ,
\end{equation}
where $\gamma_{i_s} = \sqrt{\epsilon_{i_s} \mu_{i_s} k_0^2 - k_x^2}$. The continuity of $\tilde{E}_y$ and the jump in its derivative lead to the following equations:
\begin{align}
    A_{i_s}^+ + B_{i_s}^+ &= A_{i_s}^- + B_{i_s}^-\,,\\
    A_{i_s}^+ - B_{i_s}^+ &= A_{i_s}^- - B_{i_s}^- - \frac{i\omega \mu_0 \mu_{i_s}}{\gamma_{i_s}}  I e^{i k_x x_s}\,.
\end{align}

Now it is always possible to write that $B_{i_s}^+ = r_{u} A_{i_s}^+$ and $A_{i_s}^- = r_{d} B_{i_s}^-$, whatever the surrounding of the source. The coefficients $r_d$ and $r_{u}$ are simply found in the scattering matrix corresponding to the part of the structure below (respectively above) the source as in Eq.~\eqref{eq:S_mat_side}. A simple representation of these fields and the contribution of the line source is presented in Fig.~\ref{fig:field_schematic}.
A straightforward calculation yields an expression for the coefficients representing the wave amplitudes traveling into directions upwards ($A_{i_s}^+$), respectively downwards ($B_{i_s}^-$) from the source:
\begin{align}
        A_{i_s}^+ &= \frac{-i e^{i k_x x_s}}{1 - r_{u} + (1 + r_{u})\frac{  1- r_{d} }{ 1 + r_{d}} }\frac{\omega \mu_0 \mu_{i_s}I}{ \gamma_{i_s}}\,,\\
        B_{i_s}^- &= \frac{-i e^{i k_xx_s}}{1 - r_d + (1 + r_{d})\frac{1-r_{u}}{1 + r_{u}} }\frac{\omega \mu_0 \mu_{i_s}I}{\gamma_{i_s} }\,.
\end{align}

These two coefficients represent the amplitudes of the plane waves excited by the source, and as can be clearly seen, both depend highly on the surroundings through $r_d$ and $r_{u}$, which means the surroundings do influence the source itself. This allows to compute the field above and under the source when no other incoming light is considered. Practically, we use Eq.~\eqref{eq:amplitude} with amplitude $\mathcal{E}_0(k_x)$ unity for all the values of $k_x$.

\section{Guided modes tracking}

It is extremely useful to be able to find the electromagnetic modes of a multilayered structure, whether these modes are strictly guided or leaky. Both types of modes correspond to a solution of the system of equations evoked above (see Eqs.~\eqref{eq:cont1}-\eqref{eq:cont2}) without any illumination ($B_0^- = A_{N+1}^+ =0$), meaning the determinant of the system has to be zero\cite{petit1989ondes}, \textit{i.e.,} a zero of the dispersion relation.

\begin{figure*}[ht]
    \centering
    \includegraphics[width=\linewidth]{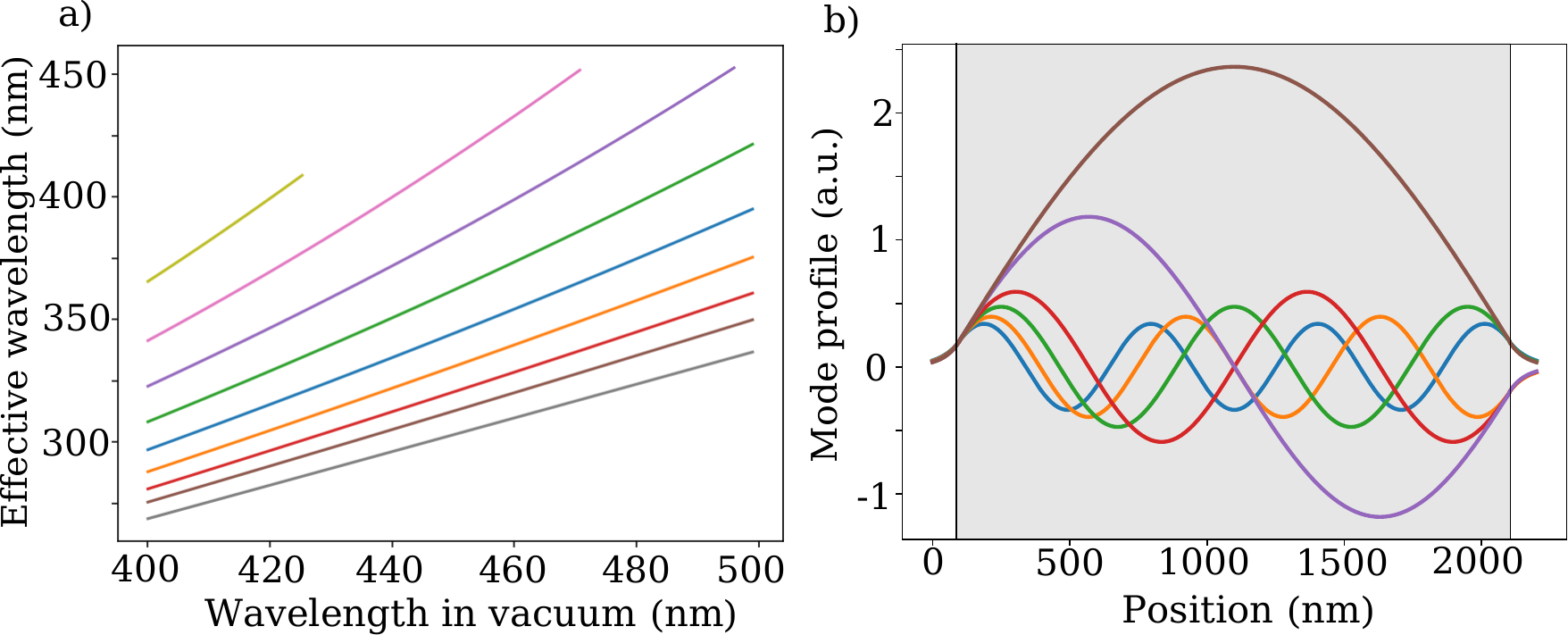}
    \caption{a) Dispersion of the guided modes of a 2000~nm thick slab of dielectric with a refractive index $n=1.5$ in air, plotted as the effective wavelength of the mode as a function of the wavelength in vacuum in a manner similar to \cite{khaywah2021large}. b) Corresponding field profiles of the 6 modes with lowest effective index.}
    \label{fig:dispersion_mode}
\end{figure*}

The dispersion relation corresponds also to a zero of the determinant of $S^{-1}$ or a pole of $S$, where $S$ is the scattering matrix of the whole structure. 
We underline, however, that this does not work for a single interface and the surface plasmon modes, for instance (the determinant of an interface scattering matrix is always equal to -1). 

A solution of the dispersion relation is also a pole of the reflection coefficient in the $k_x$ complex plane, because it corresponds to the existence of a solution or an outgoing wave without any incoming plane wave -- and thus a diverging reflection coefficient. Obtaining the guided modes in this way works in the general case, including the case of only a single interface. It is therefore the approach implemented in PyMoosh.

The solutions of the dispersion relation, and thus the poles of the reflection coefficient, can generally be found either for a real value of the wavevector $k_x$ and a complex value of the frequency $\omega$ (the imaginary part representing the typical lifetime of the mode) or for a real value of the frequency $\omega$ and a complex value of the wavevector $k_x$ (the imaginary part representing the typical propagation distance of the mode before leaking). In PyMoosh we have, so far, made the latter choice. However, the default choice for the computation of the complex square root is not adequate for finding the solution of the dispersion relation. Given the  physical meaning of the imaginary part of the poles, it is, most of the time\cite{polles2010light}, positive. The determination of the square root has to be modified to make the poles appear \cite{smith1992mode} and the default choice is to consider that the phase of $k_z$ should be comprised between $-\pi/5$ and $4\pi/5$.  This allows, in most cases, to see the poles correctly and to explore the complex plane close to them.

In order to find the guided modes, we define a minimum and a maximum effective index, defining bounds on the wavevectors.
Then we start from a set of points regularly placed on the real axis between these values for $k_x$ and run a steepest descent over $\frac{1}{|r|}$ using a finite difference scheme to estimate the gradient of the function. Steepest descent algorithms are typically employed to locate any minima, but here we leverage the fact that the minima correspond to zeroes, to determine when the algorithm should stop. Several of the initial points will converge to the same pole, and only unique poles are kept. While this technique is probably not the most efficient nor the most advanced, it has proven relatively reliable so far.

Once the modes have been found, their profile can be computed. An evanescent wave with a unity amplitude at the first interface is assumed. The rest of the mode is reconstructed using scattering matrices. Any inaccuracy in the determination of $k_x$ will result in a discontinuity at the first interface. Figure~\ref{fig:dispersion_mode}b) presents the modes found in a 2000~nm thick slab of dielectric with a refractive index $n=1.5$ in air.

It is often practical to obtain the dispersion curve, \textit{i.e.,} the values of the wavevector of a given mode for different frequencies. In PyMoosh, once the modes have been found for the largest considered frequency, they can be tracked using steepest descents while the frequency changes. An example is shown Fig. \ref{fig:dispersion_mode}a). 
We begin with the smallest wavelength so that no mode appears during the sweep, however modes do disappear when they reach a cut-off.

\section{Code design principles}

PyMoosh has been developed with the core principles of Open Science in mind. The choice of Python as the programming language plays a central role: Python packages are extremely easy to find and install through the use of the PyPI repository\cite{pypi}, Python is often called to be \textit{good glue}, thanks to its ability to interface other programming languages seamlessly. Moreover, modern repositories such as GitHub\cite{github}, being partly social networks, combined with the GNU General Public Licence, allow to maximize the accessibility and reusability of the code.
Our approach thus aligns with the principles of the FAIR data framework, which underscore the importance of making data and methods Findable, Accessible, Interoperable, and Reusable\cite{fairdata}.

\begin{figure*}[ht]
    \centering
    \includegraphics[width=\linewidth]{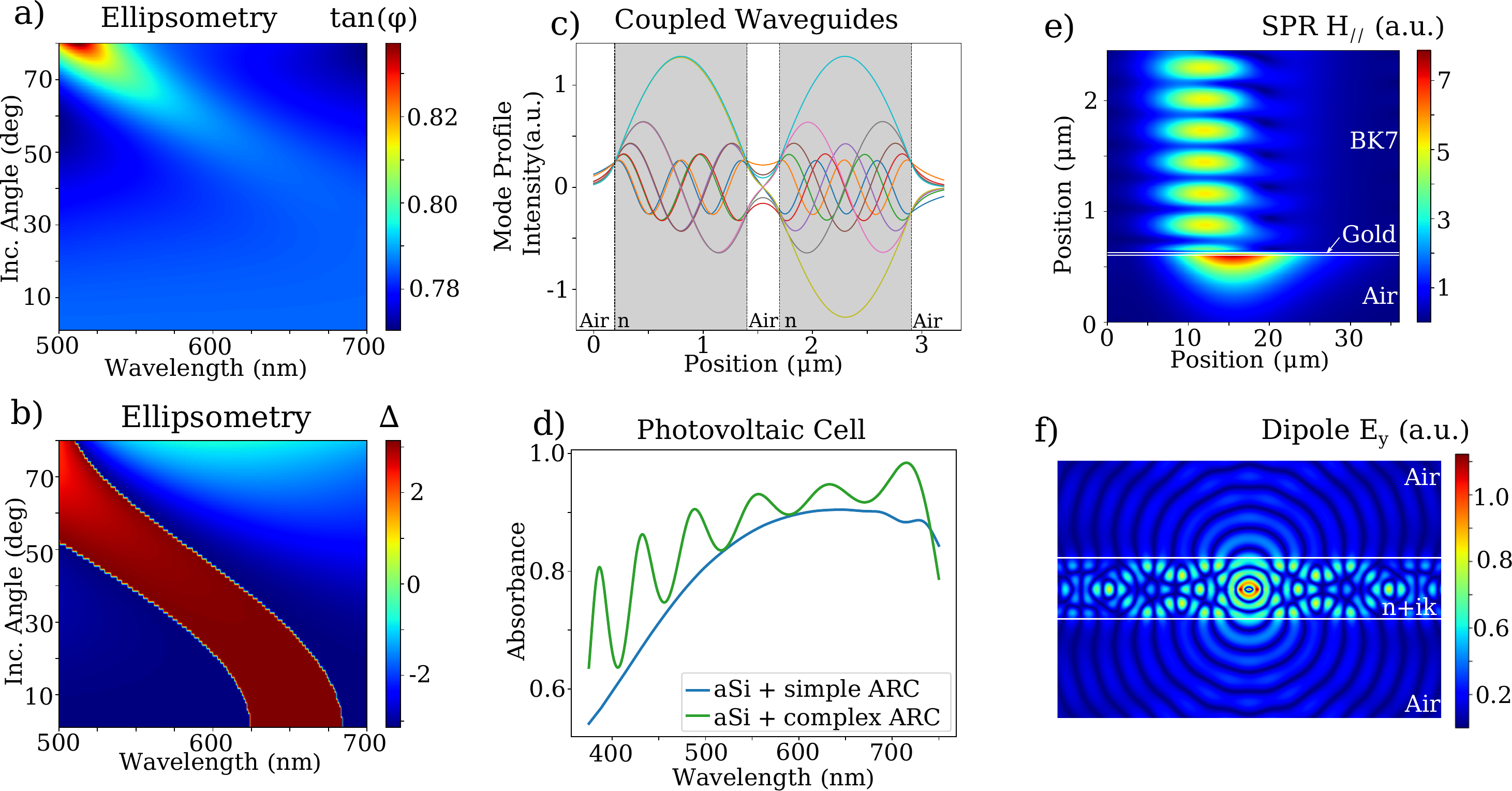}
    \caption{PyMoosh use case examples. a-b) Ellipsometry results $\frac{r_p}{r_s}=\tan(\upphi)e^{i\Delta}$ for a layer of $n=1.33$ and width 400~nm on top of a Gold substrate. c) Mode profiles for two coupled waveguides of index $n=1.5$, width 1.2~$\upmu$m and separated by 300~nm of air. d) 
    Bulk 1~$\upmu$m aSi cell absorbance in two cases. Simple Anti-Reflective Coating (ARC) : 100~nm layer of $n=1.5$ as shown in Fig.~\ref{fig:PV}b). Complex ARC: opimized 6 layer Bragg mirror and impedance matching layers above and below (see notebook). e) Surface Plasmon Resonance excitation with a prism in Kraetschmann configuration: prism made out of BK7 on top of a 55~nm layer of Gold illuminated at 45.5$^\circ$ and $\lambda=600$~nm. f) Electric field intensity of a current source encased in a layer of width 1~$\upmu$m and $\epsilon=1.4+0.1i$ at $\lambda=600$~nm.}
    \label{fig:examples}
\end{figure*}

The code architecture was chosen with simplicity of use in mind, especially for researchers and physicists in photonics. Although Python is an object-oriented language, we intentionally minimized the use of objects in order to facilitate code comprehension for non-programming experts. For convenience some parts such as materials handling are implemented as classes, but we emphasize that the part concerning materials can be used and expanded upon completely independently from the rest of the code. Finally, due to the inherent ease of implementation in Python, some interesting numerical optimizations that could be deemed fundamental (\textit{e.g.} parallel computing) are left entirely in the hand of the user.

Our code is highly versatile, which may initially disorient some users - especially those who are not familiar with Python. To address this, in addition to this tutorial, we have provided extensive documentation and illustrated various use cases using Jupyter notebooks : \href{https://github.com/AnMoreau/PyMoosh/blob/main/notebooks/Basic_tutorials/PyMoosh_Basics.ipynb}{a beginner's guide to simulation}, \href{https://github.com/AnMoreau/PyMoosh/blob/main/notebooks/Basic_tutorials/Optimizing_with_PyMoosh.ipynb}{another for optimization}, and then specific use cases: \href{https://github.com/AnMoreau/PyMoosh/blob/main/notebooks/In-depth_examples/Surface_Plasmon_Resonance.ipynb}{Surface Plasmon Resonance}, \href{https://github.com/AnMoreau/PyMoosh/blob/main/notebooks/In-depth_examples/Total_Internal_Reflection.ipynb}{Total Internal Reflection}, \href{https://github.com/AnMoreau/PyMoosh/blob/main/notebooks/In-depth_examples/Photovoltaics.ipynb}{Photovoltaics devices}, \href{https://github.com/AnMoreau/PyMoosh/blob/main/notebooks/In-depth_examples/Ellipsometry.ipynb}{Ellipsometry problems}, \href{https://github.com/AnMoreau/PyMoosh/blob/main/notebooks/In-depth_examples/Computing_Green_functions.ipynb}{Green function computation}, \href{https://github.com/AnMoreau/PyMoosh/blob/main/notebooks/In-depth_examples/Adapting_matrix_formalism.ipynb}{Use of various formalisms}, \href{https://github.com/AnMoreau/PyMoosh/blob/main/notebooks/In-depth_examples/Finding_representing_guided_modes.ipynb}{Guided Mode representation and tracking}, \href{https://github.com/AnMoreau/PyMoosh/blob/main/notebooks/In-depth_examples/Prism_couplers.ipynb}{Prism couplers},  \href{https://github.com/AnMoreau/PyMoosh/blob/main/notebooks/In-depth_examples/Photonic_Crystals_1D.ipynb}{Photonic Crystals}, and \href{https://github.com/AnMoreau/PyMoosh/blob/main/notebooks/In-depth_examples/How_materials_works.ipynb}{Managing materials}.
A sample of the results these notebooks provide is shown in Fig.~\ref{fig:examples}. Jupyter notebooks have gained significant popularity, particularly in the context of open science\cite{jupyter}, making them an accessible and user-friendly tool for exploring and utilizing our code. Finally, we illustrate possible applications for deep learning training dataset generation by an own set of open source, extensive python notebook examples, which are exhaustively discussed in a recent tutorial article \cite{khaireh2023newcomer, wiechaNewcomerGuideDeep2023}.

\section{Conclusion}

In conclusion, we have presented PyMoosh, a user-friendly Python tool for the analysis and design of photonic multilayers. PyMoosh has been developped with Open Science principles in mind, to make it genuinely Findable, Accessible, Interoperable and Reusable (FAIR data approach).  With its comprehensive set of features, numerical stability, readiness for optimization and deep learning problems and its extensive set of examples, we hope our platform will be adopted by the community at large to build upon. Already utilized in teaching and research, PyMoosh offers a versatile platform for studying the rich physics of multilayered structures, particularly those incorporating metallic or metallic-like layers. In a subsequent publication, 
we aim to demonstrate how PyMoosh can be seamlessly integrated with advanced optimization algorithms and optimization libraries and that its features allow for an efficient exploration of inverse design problems\cite{bennet2023illustrated,khaireh2023newcomer}. Future plans include incorporating more advanced material descriptions, such as spatial dispersion in metals\cite{melnyk1970theory,benedicto2015numerical}, to further expand the capabilities of PyMoosh. We are dedicated to continuously improving PyMoosh and welcome any suggestions and contributions from the community.

\section*{Acknowledgements}
A.M. is an Academy CAP 20-25 chair holder. He acknowledges the support received from the Agence Nationale de la Recherche of the French government through the program Investissements d'Avenir (16-IDEX-0001 CAP 20-25). This work was supported by the International Research Center "Innovation Transportation and Production Systems" of the Clermont-Ferrand I-SITE CAP 20-25. P.R.W. acknowledges the support of the French Agence Nationale de la Recherche (ANR) under grant ANR-22-CE24-0002 (project NAINOS), and from the Toulouse high performance computing facility CALMIP (grant p20010).

\bibliography{apssamp}

\begin{thebibliography}{67}%
\makeatletter
\providecommand \@ifxundefined [1]{%
 \@ifx{#1\undefined}
}%
\providecommand \@ifnum [1]{%
 \ifnum #1\expandafter \@firstoftwo
 \else \expandafter \@secondoftwo
 \fi
}%
\providecommand \@ifx [1]{%
 \ifx #1\expandafter \@firstoftwo
 \else \expandafter \@secondoftwo
 \fi
}%
\providecommand \natexlab [1]{#1}%
\providecommand \enquote  [1]{``#1''}%
\providecommand \bibnamefont  [1]{#1}%
\providecommand \bibfnamefont [1]{#1}%
\providecommand \citenamefont [1]{#1}%
\providecommand \href@noop [0]{\@secondoftwo}%
\providecommand \href [0]{\begingroup \@sanitize@url \@href}%
\providecommand \@href[1]{\@@startlink{#1}\@@href}%
\providecommand \@@href[1]{\endgroup#1\@@endlink}%
\providecommand \@sanitize@url [0]{\catcode `\\12\catcode `\$12\catcode
  `\&12\catcode `\#12\catcode `\^12\catcode `\_12\catcode `\%12\relax}%
\providecommand \@@startlink[1]{}%
\providecommand \@@endlink[0]{}%
\providecommand \url  [0]{\begingroup\@sanitize@url \@url }%
\providecommand \@url [1]{\endgroup\@href {#1}{\urlprefix }}%
\providecommand \urlprefix  [0]{URL }%
\providecommand \Eprint [0]{\href }%
\providecommand \doibase [0]{https://doi.org/}%
\providecommand \selectlanguage [0]{\@gobble}%
\providecommand \bibinfo  [0]{\@secondoftwo}%
\providecommand \bibfield  [0]{\@secondoftwo}%
\providecommand \translation [1]{[#1]}%
\providecommand \BibitemOpen [0]{}%
\providecommand \bibitemStop [0]{}%
\providecommand \bibitemNoStop [0]{.\EOS\space}%
\providecommand \EOS [0]{\spacefactor3000\relax}%
\providecommand \BibitemShut  [1]{\csname bibitem#1\endcsname}%
\let\auto@bib@innerbib\@empty
\bibitem [{\citenamefont {Strutt}(1912)}]{strutt1912propagation}%
  \BibitemOpen
  \bibfield  {author} {\bibinfo {author} {\bibfnamefont {J.~W.}\ \bibnamefont
  {Strutt}},\ }\href@noop {} {\bibfield  {journal} {\bibinfo  {journal}
  {Proceedings of the Royal Society of London. Series A, Containing Papers of a
  Mathematical and Physical Character}\ }\textbf {\bibinfo {volume} {86}},\
  \bibinfo {pages} {207} (\bibinfo {year} {1912})}\BibitemShut {NoStop}%
\bibitem [{\citenamefont {Strutt}(1917)}]{strutt1917reflection}%
  \BibitemOpen
  \bibfield  {author} {\bibinfo {author} {\bibfnamefont {J.~W.}\ \bibnamefont
  {Strutt}},\ }\href@noop {} {\bibfield  {journal} {\bibinfo  {journal}
  {Proceedings of the Royal Society of London. Series A, Containing Papers of a
  Mathematical and Physical Character}\ }\textbf {\bibinfo {volume} {93}},\
  \bibinfo {pages} {565} (\bibinfo {year} {1917})}\BibitemShut {NoStop}%
\bibitem [{\citenamefont {Macleod}(2012)}]{macleod2012quarterwave}%
  \BibitemOpen
  \bibfield  {author} {\bibinfo {author} {\bibfnamefont {A.}~\bibnamefont
  {Macleod}},\ }\href@noop {} {\bibfield  {journal} {\bibinfo  {journal}
  {Bulletin, Society of Vacuum Coaters, Issue Summer}\ ,\ \bibinfo {pages}
  {22}} (\bibinfo {year} {2012})}\BibitemShut {NoStop}%
\bibitem [{\citenamefont {Abel{\`e}s}(1950)}]{Abeles1950theorie}%
  \BibitemOpen
  \bibfield  {author} {\bibinfo {author} {\bibfnamefont {F.}~\bibnamefont
  {Abel{\`e}s}},\ }\href@noop {} {\bibfield  {journal} {\bibinfo  {journal}
  {Journal de Physique et le Radium}\ }\textbf {\bibinfo {volume} {11}},\
  \bibinfo {pages} {307} (\bibinfo {year} {1950})}\BibitemShut {NoStop}%
\bibitem [{\citenamefont {Yeh}\ \emph {et~al.}(1977)\citenamefont {Yeh},
  \citenamefont {Yariv},\ and\ \citenamefont {Hong}}]{yeh1977electromagnetic}%
  \BibitemOpen
  \bibfield  {author} {\bibinfo {author} {\bibfnamefont {P.}~\bibnamefont
  {Yeh}}, \bibinfo {author} {\bibfnamefont {A.}~\bibnamefont {Yariv}},\ and\
  \bibinfo {author} {\bibfnamefont {C.-S.}\ \bibnamefont {Hong}},\ }\href@noop
  {} {\bibfield  {journal} {\bibinfo  {journal} {JOSA}\ }\textbf {\bibinfo
  {volume} {67}},\ \bibinfo {pages} {423} (\bibinfo {year} {1977})}\BibitemShut
  {NoStop}%
\bibitem [{\citenamefont {Yariv}\ and\ \citenamefont
  {Yeh}(1977)}]{yariv1977electromagnetic}%
  \BibitemOpen
  \bibfield  {author} {\bibinfo {author} {\bibfnamefont {A.}~\bibnamefont
  {Yariv}}\ and\ \bibinfo {author} {\bibfnamefont {P.}~\bibnamefont {Yeh}},\
  }\href@noop {} {\bibfield  {journal} {\bibinfo  {journal} {JOSA}\ }\textbf
  {\bibinfo {volume} {67}},\ \bibinfo {pages} {438} (\bibinfo {year}
  {1977})}\BibitemShut {NoStop}%
\bibitem [{\citenamefont {Yeh}\ \emph {et~al.}(1978)\citenamefont {Yeh},
  \citenamefont {Yariv},\ and\ \citenamefont {Cho}}]{yeh1978optical}%
  \BibitemOpen
  \bibfield  {author} {\bibinfo {author} {\bibfnamefont {P.}~\bibnamefont
  {Yeh}}, \bibinfo {author} {\bibfnamefont {A.}~\bibnamefont {Yariv}},\ and\
  \bibinfo {author} {\bibfnamefont {A.~Y.}\ \bibnamefont {Cho}},\ }\href@noop
  {} {\bibfield  {journal} {\bibinfo  {journal} {Applied Physics Letters}\
  }\textbf {\bibinfo {volume} {32}},\ \bibinfo {pages} {104} (\bibinfo {year}
  {1978})}\BibitemShut {NoStop}%
\bibitem [{\citenamefont {Yeh}(1980)}]{yeh1980optics}%
  \BibitemOpen
  \bibfield  {author} {\bibinfo {author} {\bibfnamefont {P.}~\bibnamefont
  {Yeh}},\ }\href@noop {} {\bibfield  {journal} {\bibinfo  {journal} {Surface
  Science}\ }\textbf {\bibinfo {volume} {96}},\ \bibinfo {pages} {41} (\bibinfo
  {year} {1980})}\BibitemShut {NoStop}%
\bibitem [{\citenamefont {Born}\ and\ \citenamefont
  {Wolf}(2013)}]{born2013principles}%
  \BibitemOpen
  \bibfield  {author} {\bibinfo {author} {\bibfnamefont {M.}~\bibnamefont
  {Born}}\ and\ \bibinfo {author} {\bibfnamefont {E.}~\bibnamefont {Wolf}},\
  }\href@noop {} {\emph {\bibinfo {title} {Principles of optics:
  electromagnetic theory of propagation, interference and diffraction of
  light}}}\ (\bibinfo  {publisher} {Elsevier},\ \bibinfo {year}
  {2013})\BibitemShut {NoStop}%
\bibitem [{\citenamefont {Yeh}(2006)}]{yeh2006optical}%
  \BibitemOpen
  \bibfield  {author} {\bibinfo {author} {\bibfnamefont {P.}~\bibnamefont
  {Yeh}},\ }\href@noop {} {\emph {\bibinfo {title} {Optical Waves in Layered
  Media}}}\ (\bibinfo  {publisher} {Wiley Online Library},\ \bibinfo {year}
  {2006})\BibitemShut {NoStop}%
\bibitem [{\citenamefont {Macleod}(2017)}]{macleod2017thin}%
  \BibitemOpen
  \bibfield  {author} {\bibinfo {author} {\bibfnamefont {H.~A.}\ \bibnamefont
  {Macleod}},\ }\href@noop {} {\emph {\bibinfo {title} {Thin-film optical
  filters}}}\ (\bibinfo  {publisher} {CRC press},\ \bibinfo {year}
  {2017})\BibitemShut {NoStop}%
\bibitem [{\citenamefont {Baumeister}(1958)}]{baumeister_design_1958}%
  \BibitemOpen
  \bibfield  {author} {\bibinfo {author} {\bibfnamefont {P.}~\bibnamefont
  {Baumeister}},\ }\href@noop {} {\bibfield  {journal} {\bibinfo  {journal}
  {JOSA}\ }\textbf {\bibinfo {volume} {48}} (\bibinfo {year}
  {1958})}\BibitemShut {NoStop}%
\bibitem [{\citenamefont {Thelen}(1966)}]{thelen_equivalent_1966}%
  \BibitemOpen
  \bibfield  {author} {\bibinfo {author} {\bibfnamefont {A.}~\bibnamefont
  {Thelen}},\ }\bibfield  {journal} {\bibinfo  {journal} {JOSA}\ }\textbf
  {\bibinfo {volume} {56}},\ \href {https://doi.org/10.1364/JOSA.56.001533}
  {10.1364/JOSA.56.001533} (\bibinfo {year} {1966})\BibitemShut {NoStop}%
\bibitem [{\citenamefont {Thelen}(1971)}]{thelen_design_1971}%
  \BibitemOpen
  \bibfield  {author} {\bibinfo {author} {\bibfnamefont {A.}~\bibnamefont
  {Thelen}},\ }\bibfield  {journal} {\bibinfo  {journal} {JOSA}\ }\textbf
  {\bibinfo {volume} {61}},\ \href {https://doi.org/10.1364/JOSA.61.000365}
  {10.1364/JOSA.61.000365} (\bibinfo {year} {1971})\BibitemShut {NoStop}%
\bibitem [{\citenamefont {Apfel}(1977)}]{apfel_optical_1977}%
  \BibitemOpen
  \bibfield  {author} {\bibinfo {author} {\bibfnamefont {J.~H.}\ \bibnamefont
  {Apfel}},\ }\bibfield  {journal} {\bibinfo  {journal} {Applied Optics}\
  }\textbf {\bibinfo {volume} {16}},\ \href
  {https://doi.org/10.1364/AO.16.001880} {10.1364/AO.16.001880} (\bibinfo
  {year} {1977})\BibitemShut {NoStop}%
\bibitem [{\citenamefont {Dobrowolski}\ and\ \citenamefont
  {Lowe}(1978)}]{dobrowolski_optical_1978}%
  \BibitemOpen
  \bibfield  {author} {\bibinfo {author} {\bibfnamefont {J.}~\bibnamefont
  {Dobrowolski}}\ and\ \bibinfo {author} {\bibfnamefont {D.}~\bibnamefont
  {Lowe}},\ }\href@noop {} {\bibfield  {journal} {\bibinfo  {journal} {Applied
  Optics}\ }\textbf {\bibinfo {volume} {17}} (\bibinfo {year}
  {1978})}\BibitemShut {NoStop}%
\bibitem [{\citenamefont {Tikhonravov}\ and\ \citenamefont
  {Trubetskov}(1994)}]{tikhonravov1994development}%
  \BibitemOpen
  \bibfield  {author} {\bibinfo {author} {\bibfnamefont {A.~V.}\ \bibnamefont
  {Tikhonravov}}\ and\ \bibinfo {author} {\bibfnamefont {M.~K.}\ \bibnamefont
  {Trubetskov}},\ }in\ \href@noop {} {\emph {\bibinfo {booktitle} {Optical
  Interference Coatings}}},\ Vol.\ \bibinfo {volume} {2253}\ (\bibinfo
  {organization} {SPIE},\ \bibinfo {year} {1994})\ pp.\ \bibinfo {pages}
  {10--20}\BibitemShut {NoStop}%
\bibitem [{\citenamefont {Poitras}\ \emph {et~al.}(2017)\citenamefont
  {Poitras}, \citenamefont {Li}, \citenamefont {Jacobson},\ and\ \citenamefont
  {Cooksey}}]{poitras_manufacturing_2017}%
  \BibitemOpen
  \bibfield  {author} {\bibinfo {author} {\bibfnamefont {D.}~\bibnamefont
  {Poitras}}, \bibinfo {author} {\bibfnamefont {L.}~\bibnamefont {Li}},
  \bibinfo {author} {\bibfnamefont {M.}~\bibnamefont {Jacobson}},\ and\
  \bibinfo {author} {\bibfnamefont {C.}~\bibnamefont {Cooksey}},\ }\bibfield
  {journal} {\bibinfo  {journal} {Applied Optics}\ }\textbf {\bibinfo {volume}
  {56}},\ \href {https://doi.org/10.1364/AO.56.0000C1} {10.1364/AO.56.0000C1}
  (\bibinfo {year} {2017})\BibitemShut {NoStop}%
\bibitem [{\citenamefont {Kruschwitz}\ \emph {et~al.}(2017)\citenamefont
  {Kruschwitz}, \citenamefont {Pervak}, \citenamefont {Keck}, \citenamefont
  {Bolshakov}, \citenamefont {Gerig}, \citenamefont {Lemarchand}, \citenamefont
  {Sato}, \citenamefont {Southwell}, \citenamefont {Sugiura}, \citenamefont
  {Trubetskov},\ and\ \citenamefont {Yuan}}]{kruschwitz_optical_2017}%
  \BibitemOpen
  \bibfield  {author} {\bibinfo {author} {\bibfnamefont {J.~D.~T.}\
  \bibnamefont {Kruschwitz}}, \bibinfo {author} {\bibfnamefont
  {V.}~\bibnamefont {Pervak}}, \bibinfo {author} {\bibfnamefont
  {J.}~\bibnamefont {Keck}}, \bibinfo {author} {\bibfnamefont {I.}~\bibnamefont
  {Bolshakov}}, \bibinfo {author} {\bibfnamefont {Z.}~\bibnamefont {Gerig}},
  \bibinfo {author} {\bibfnamefont {F.}~\bibnamefont {Lemarchand}}, \bibinfo
  {author} {\bibfnamefont {K.}~\bibnamefont {Sato}}, \bibinfo {author}
  {\bibfnamefont {W.}~\bibnamefont {Southwell}}, \bibinfo {author}
  {\bibfnamefont {M.}~\bibnamefont {Sugiura}}, \bibinfo {author} {\bibfnamefont
  {M.}~\bibnamefont {Trubetskov}},\ and\ \bibinfo {author} {\bibfnamefont
  {W.}~\bibnamefont {Yuan}},\ }\bibfield  {journal} {\bibinfo  {journal}
  {Applied Optics}\ }\textbf {\bibinfo {volume} {56}},\ \href
  {https://doi.org/10.1364/AO.56.00C151} {10.1364/AO.56.00C151} (\bibinfo
  {year} {2017})\BibitemShut {NoStop}%
\bibitem [{\citenamefont {Kruschwitz}\ \emph {et~al.}(2019)\citenamefont
  {Kruschwitz}, \citenamefont {Pervak},\ and\ \citenamefont
  {Keck}}]{Kruschwitz:19}%
  \BibitemOpen
  \bibfield  {author} {\bibinfo {author} {\bibfnamefont {J.~D.~T.}\
  \bibnamefont {Kruschwitz}}, \bibinfo {author} {\bibfnamefont
  {V.}~\bibnamefont {Pervak}},\ and\ \bibinfo {author} {\bibfnamefont
  {J.}~\bibnamefont {Keck}},\ }in\ \href
  {https://doi.org/10.1364/OIC.2019.TC.1} {\emph {\bibinfo {booktitle} {Optical
  Interference Coatings Conference (OIC) 2019}}}\ (\bibinfo  {publisher}
  {Optica Publishing Group},\ \bibinfo {year} {2019})\ p.\ \bibinfo {pages}
  {TC.1}\BibitemShut {NoStop}%
\bibitem [{\citenamefont {Bockov{\'a}}\ \emph {et~al.}(2019)\citenamefont
  {Bockov{\'a}}, \citenamefont {Slab{\`y}}, \citenamefont {{\v{S}}pringer},\
  and\ \citenamefont {Homola}}]{bockova2019advances}%
  \BibitemOpen
  \bibfield  {author} {\bibinfo {author} {\bibfnamefont {M.}~\bibnamefont
  {Bockov{\'a}}}, \bibinfo {author} {\bibfnamefont {J.}~\bibnamefont
  {Slab{\`y}}}, \bibinfo {author} {\bibfnamefont {T.}~\bibnamefont
  {{\v{S}}pringer}},\ and\ \bibinfo {author} {\bibfnamefont {J.}~\bibnamefont
  {Homola}},\ }\href@noop {} {\bibfield  {journal} {\bibinfo  {journal} {Annual
  Review of Analytical Chemistry}\ }\textbf {\bibinfo {volume} {12}},\ \bibinfo
  {pages} {151} (\bibinfo {year} {2019})}\BibitemShut {NoStop}%
\bibitem [{\citenamefont {Raut}\ \emph {et~al.}(2011)\citenamefont {Raut},
  \citenamefont {Ganesh}, \citenamefont {Nair},\ and\ \citenamefont
  {Ramakrishna}}]{raut2011anti}%
  \BibitemOpen
  \bibfield  {author} {\bibinfo {author} {\bibfnamefont {H.~K.}\ \bibnamefont
  {Raut}}, \bibinfo {author} {\bibfnamefont {V.~A.}\ \bibnamefont {Ganesh}},
  \bibinfo {author} {\bibfnamefont {A.~S.}\ \bibnamefont {Nair}},\ and\
  \bibinfo {author} {\bibfnamefont {S.}~\bibnamefont {Ramakrishna}},\
  }\href@noop {} {\bibfield  {journal} {\bibinfo  {journal} {Energy \&
  Environmental Science}\ }\textbf {\bibinfo {volume} {4}},\ \bibinfo {pages}
  {3779} (\bibinfo {year} {2011})}\BibitemShut {NoStop}%
\bibitem [{\citenamefont {Bozhevolnyi}\ and\ \citenamefont
  {S{\o}ndergaard}(2007)}]{bozhevolnyi2007general}%
  \BibitemOpen
  \bibfield  {author} {\bibinfo {author} {\bibfnamefont {S.~I.}\ \bibnamefont
  {Bozhevolnyi}}\ and\ \bibinfo {author} {\bibfnamefont {T.}~\bibnamefont
  {S{\o}ndergaard}},\ }\href@noop {} {\bibfield  {journal} {\bibinfo  {journal}
  {Optics express}\ }\textbf {\bibinfo {volume} {15}},\ \bibinfo {pages}
  {10869} (\bibinfo {year} {2007})}\BibitemShut {NoStop}%
\bibitem [{\citenamefont {Shekhar}\ \emph {et~al.}(2014)\citenamefont
  {Shekhar}, \citenamefont {Atkinson},\ and\ \citenamefont
  {Jacob}}]{shekhar2014hyperbolic}%
  \BibitemOpen
  \bibfield  {author} {\bibinfo {author} {\bibfnamefont {P.}~\bibnamefont
  {Shekhar}}, \bibinfo {author} {\bibfnamefont {J.}~\bibnamefont {Atkinson}},\
  and\ \bibinfo {author} {\bibfnamefont {Z.}~\bibnamefont {Jacob}},\
  }\href@noop {} {\bibfield  {journal} {\bibinfo  {journal} {Nano convergence}\
  }\textbf {\bibinfo {volume} {1}},\ \bibinfo {pages} {1} (\bibinfo {year}
  {2014})}\BibitemShut {NoStop}%
\bibitem [{\citenamefont {Poll{\`e}s}\ \emph {et~al.}(2016)\citenamefont
  {Poll{\`e}s}, \citenamefont {Mihailovic}, \citenamefont {Centeno},\ and\
  \citenamefont {Moreau}}]{polles2016leveraging}%
  \BibitemOpen
  \bibfield  {author} {\bibinfo {author} {\bibfnamefont {R.}~\bibnamefont
  {Poll{\`e}s}}, \bibinfo {author} {\bibfnamefont {M.}~\bibnamefont
  {Mihailovic}}, \bibinfo {author} {\bibfnamefont {E.}~\bibnamefont
  {Centeno}},\ and\ \bibinfo {author} {\bibfnamefont {A.}~\bibnamefont
  {Moreau}},\ }\href@noop {} {\bibfield  {journal} {\bibinfo  {journal}
  {Physical Review A}\ }\textbf {\bibinfo {volume} {94}},\ \bibinfo {pages}
  {063808} (\bibinfo {year} {2016})}\BibitemShut {NoStop}%
\bibitem [{\citenamefont {Katsidis}\ and\ \citenamefont
  {Siapkas}(2002)}]{katsidisGeneralTransfermatrixMethod2002}%
  \BibitemOpen
  \bibfield  {author} {\bibinfo {author} {\bibfnamefont {C.~C.}\ \bibnamefont
  {Katsidis}}\ and\ \bibinfo {author} {\bibfnamefont {D.~I.}\ \bibnamefont
  {Siapkas}},\ }\href {https://doi.org/10.1364/AO.41.003978} {\bibfield
  {journal} {\bibinfo  {journal} {Applied Optics}\ }\textbf {\bibinfo {volume}
  {41}},\ \bibinfo {pages} {3978} (\bibinfo {year} {2002})}\BibitemShut
  {NoStop}%
\bibitem [{\citenamefont {Luce}\ \emph {et~al.}(2022)\citenamefont {Luce},
  \citenamefont {Mahdavi}, \citenamefont {Marquardt},\ and\ \citenamefont
  {Wankerl}}]{luceTMMFastTransferMatrix2022}%
  \BibitemOpen
  \bibfield  {author} {\bibinfo {author} {\bibfnamefont {A.}~\bibnamefont
  {Luce}}, \bibinfo {author} {\bibfnamefont {A.}~\bibnamefont {Mahdavi}},
  \bibinfo {author} {\bibfnamefont {F.}~\bibnamefont {Marquardt}},\ and\
  \bibinfo {author} {\bibfnamefont {H.}~\bibnamefont {Wankerl}},\ }\href
  {https://doi.org/10.1364/JOSAA.450928} {\bibfield  {journal} {\bibinfo
  {journal} {JOSA A}\ }\textbf {\bibinfo {volume} {39}},\ \bibinfo {pages}
  {1007} (\bibinfo {year} {2022})}\BibitemShut {NoStop}%
\bibitem [{\citenamefont {Bay}\ \emph {et~al.}(2022)\citenamefont {Bay},
  \citenamefont {Vignolini},\ and\ \citenamefont
  {Vynck}}]{bayPyLlamaStableVersatile2022}%
  \BibitemOpen
  \bibfield  {author} {\bibinfo {author} {\bibfnamefont {M.~M.}\ \bibnamefont
  {Bay}}, \bibinfo {author} {\bibfnamefont {S.}~\bibnamefont {Vignolini}},\
  and\ \bibinfo {author} {\bibfnamefont {K.}~\bibnamefont {Vynck}},\ }\href
  {https://doi.org/10.1016/j.cpc.2021.108256} {\bibfield  {journal} {\bibinfo
  {journal} {Computer Physics Communications}\ }\textbf {\bibinfo {volume}
  {273}},\ \bibinfo {pages} {108256} (\bibinfo {year} {2022})}\BibitemShut
  {NoStop}%
\bibitem [{\citenamefont {Larouche}\ and\ \citenamefont
  {Martinu}(2008)}]{larouche2008openfilters}%
  \BibitemOpen
  \bibfield  {author} {\bibinfo {author} {\bibfnamefont {S.}~\bibnamefont
  {Larouche}}\ and\ \bibinfo {author} {\bibfnamefont {L.}~\bibnamefont
  {Martinu}},\ }\href@noop {} {\bibfield  {journal} {\bibinfo  {journal}
  {Applied optics}\ }\textbf {\bibinfo {volume} {47}},\ \bibinfo {pages} {C219}
  (\bibinfo {year} {2008})}\BibitemShut {NoStop}%
\bibitem [{\citenamefont {Costa}\ \emph {et~al.}(2019)\citenamefont {Costa},
  \citenamefont {Rodrigues},\ and\ \citenamefont {Pereira}}]{costa2019sim}%
  \BibitemOpen
  \bibfield  {author} {\bibinfo {author} {\bibfnamefont {E.~B.}\ \bibnamefont
  {Costa}}, \bibinfo {author} {\bibfnamefont {E.~P.}\ \bibnamefont
  {Rodrigues}},\ and\ \bibinfo {author} {\bibfnamefont {H.~A.}\ \bibnamefont
  {Pereira}},\ }\href@noop {} {\bibfield  {journal} {\bibinfo  {journal}
  {Plasmonics}\ }\textbf {\bibinfo {volume} {14}},\ \bibinfo {pages} {1699}
  (\bibinfo {year} {2019})}\BibitemShut {NoStop}%
\bibitem [{\citenamefont {Barry}\ \emph {et~al.}(2020)\citenamefont {Barry},
  \citenamefont {Berthier}, \citenamefont {Wilts}, \citenamefont {Cambourieux},
  \citenamefont {Bennet}, \citenamefont {Poll{\`e}s}, \citenamefont {Teytaud},
  \citenamefont {Centeno}, \citenamefont {Biais},\ and\ \citenamefont
  {Moreau}}]{barry2020evolutionary}%
  \BibitemOpen
  \bibfield  {author} {\bibinfo {author} {\bibfnamefont {M.~A.}\ \bibnamefont
  {Barry}}, \bibinfo {author} {\bibfnamefont {V.}~\bibnamefont {Berthier}},
  \bibinfo {author} {\bibfnamefont {B.~D.}\ \bibnamefont {Wilts}}, \bibinfo
  {author} {\bibfnamefont {M.-C.}\ \bibnamefont {Cambourieux}}, \bibinfo
  {author} {\bibfnamefont {P.}~\bibnamefont {Bennet}}, \bibinfo {author}
  {\bibfnamefont {R.}~\bibnamefont {Poll{\`e}s}}, \bibinfo {author}
  {\bibfnamefont {O.}~\bibnamefont {Teytaud}}, \bibinfo {author} {\bibfnamefont
  {E.}~\bibnamefont {Centeno}}, \bibinfo {author} {\bibfnamefont
  {N.}~\bibnamefont {Biais}},\ and\ \bibinfo {author} {\bibfnamefont
  {A.}~\bibnamefont {Moreau}},\ }\href@noop {} {\bibfield  {journal} {\bibinfo
  {journal} {Scientific reports}\ }\textbf {\bibinfo {volume} {10}},\ \bibinfo
  {pages} {12024} (\bibinfo {year} {2020})}\BibitemShut {NoStop}%
\bibitem [{\citenamefont {Wankerl}\ \emph {et~al.}(2022)\citenamefont
  {Wankerl}, \citenamefont {Wiesmann}, \citenamefont {Kreiner}, \citenamefont
  {Butendeich}, \citenamefont {Luce}, \citenamefont {Sobczyk}, \citenamefont
  {Stern},\ and\ \citenamefont {Lang}}]{wankerl2022directional}%
  \BibitemOpen
  \bibfield  {author} {\bibinfo {author} {\bibfnamefont {H.}~\bibnamefont
  {Wankerl}}, \bibinfo {author} {\bibfnamefont {C.}~\bibnamefont {Wiesmann}},
  \bibinfo {author} {\bibfnamefont {L.}~\bibnamefont {Kreiner}}, \bibinfo
  {author} {\bibfnamefont {R.}~\bibnamefont {Butendeich}}, \bibinfo {author}
  {\bibfnamefont {A.}~\bibnamefont {Luce}}, \bibinfo {author} {\bibfnamefont
  {S.}~\bibnamefont {Sobczyk}}, \bibinfo {author} {\bibfnamefont {M.~L.}\
  \bibnamefont {Stern}},\ and\ \bibinfo {author} {\bibfnamefont {E.~W.}\
  \bibnamefont {Lang}},\ }\href@noop {} {\bibfield  {journal} {\bibinfo
  {journal} {Scientific Reports}\ }\textbf {\bibinfo {volume} {12}},\ \bibinfo
  {pages} {5226} (\bibinfo {year} {2022})}\BibitemShut {NoStop}%
\bibitem [{\citenamefont {Liu}\ \emph {et~al.}(2018)\citenamefont {Liu},
  \citenamefont {Zhu}, \citenamefont {Rodrigues}, \citenamefont {Lee},\ and\
  \citenamefont {Cai}}]{liuGenerativeModelInverse2018}%
  \BibitemOpen
  \bibfield  {author} {\bibinfo {author} {\bibfnamefont {Z.}~\bibnamefont
  {Liu}}, \bibinfo {author} {\bibfnamefont {D.}~\bibnamefont {Zhu}}, \bibinfo
  {author} {\bibfnamefont {S.~P.}\ \bibnamefont {Rodrigues}}, \bibinfo {author}
  {\bibfnamefont {K.-T.}\ \bibnamefont {Lee}},\ and\ \bibinfo {author}
  {\bibfnamefont {W.}~\bibnamefont {Cai}},\ }\href
  {https://doi.org/10.1021/acs.nanolett.8b03171} {\bibfield  {journal}
  {\bibinfo  {journal} {Nano Letters}\ }\textbf {\bibinfo {volume} {18}},\
  \bibinfo {pages} {6570} (\bibinfo {year} {2018})}\BibitemShut {NoStop}%
\bibitem [{\citenamefont {Unni}\ \emph {et~al.}(2020)\citenamefont {Unni},
  \citenamefont {Yao},\ and\ \citenamefont
  {Zheng}}]{unniDeepConvolutionalMixture2020}%
  \BibitemOpen
  \bibfield  {author} {\bibinfo {author} {\bibfnamefont {R.}~\bibnamefont
  {Unni}}, \bibinfo {author} {\bibfnamefont {K.}~\bibnamefont {Yao}},\ and\
  \bibinfo {author} {\bibfnamefont {Y.}~\bibnamefont {Zheng}},\ }\bibfield
  {journal} {\bibinfo  {journal} {ACS Photonics}\ }\textbf {\bibinfo {volume}
  {7}},\ \href {https://doi.org/10.1021/acsphotonics.0c00630}
  {10.1021/acsphotonics.0c00630} (\bibinfo {year} {2020})\BibitemShut {NoStop}%
\bibitem [{\citenamefont {Dai}\ \emph {et~al.}(2021)\citenamefont {Dai},
  \citenamefont {Wang}, \citenamefont {Hu}, \citenamefont {de~Groot},
  \citenamefont {Muskens}, \citenamefont {Duan},\ and\ \citenamefont
  {Huang}}]{daiAccurateInverseDesign2021}%
  \BibitemOpen
  \bibfield  {author} {\bibinfo {author} {\bibfnamefont {P.}~\bibnamefont
  {Dai}}, \bibinfo {author} {\bibfnamefont {Y.}~\bibnamefont {Wang}}, \bibinfo
  {author} {\bibfnamefont {Y.}~\bibnamefont {Hu}}, \bibinfo {author}
  {\bibfnamefont {C.~H.}\ \bibnamefont {de~Groot}}, \bibinfo {author}
  {\bibfnamefont {O.}~\bibnamefont {Muskens}}, \bibinfo {author} {\bibfnamefont
  {H.}~\bibnamefont {Duan}},\ and\ \bibinfo {author} {\bibfnamefont
  {R.}~\bibnamefont {Huang}},\ }\href {https://doi.org/10.1364/PRJ.415141}
  {\bibfield  {journal} {\bibinfo  {journal} {Photonics Research}\ }\textbf
  {\bibinfo {volume} {9}},\ \bibinfo {pages} {B236} (\bibinfo {year}
  {2021})}\BibitemShut {NoStop}%
\bibitem [{\citenamefont {Dai}\ \emph {et~al.}(2022)\citenamefont {Dai},
  \citenamefont {Sun}, \citenamefont {Yan}, \citenamefont {Muskens},
  \citenamefont {de~Groot}, \citenamefont {Zhu}, \citenamefont {Hu},
  \citenamefont {Duan},\ and\ \citenamefont
  {Huang}}]{daiInverseDesignStructural2022}%
  \BibitemOpen
  \bibfield  {author} {\bibinfo {author} {\bibfnamefont {P.}~\bibnamefont
  {Dai}}, \bibinfo {author} {\bibfnamefont {K.}~\bibnamefont {Sun}}, \bibinfo
  {author} {\bibfnamefont {X.}~\bibnamefont {Yan}}, \bibinfo {author}
  {\bibfnamefont {O.~L.}\ \bibnamefont {Muskens}}, \bibinfo {author}
  {\bibfnamefont {C.~H.~K.}\ \bibnamefont {de~Groot}}, \bibinfo {author}
  {\bibfnamefont {X.}~\bibnamefont {Zhu}}, \bibinfo {author} {\bibfnamefont
  {Y.}~\bibnamefont {Hu}}, \bibinfo {author} {\bibfnamefont {H.}~\bibnamefont
  {Duan}},\ and\ \bibinfo {author} {\bibfnamefont {R.}~\bibnamefont {Huang}},\
  }\href {https://doi.org/10.1515/nanoph-2022-0095} {\bibfield  {journal}
  {\bibinfo  {journal} {Nanophotonics}\ }\textbf {\bibinfo {volume} {11}},\
  \bibinfo {pages} {3057} (\bibinfo {year} {2022})}\BibitemShut {NoStop}%
\bibitem [{\citenamefont {Wang}\ \emph {et~al.}(2022)\citenamefont {Wang},
  \citenamefont {Lin}, \citenamefont {Zhang}, \citenamefont {Wu},\ and\
  \citenamefont {Huang}}]{wangEllipsoNetDeeplearningenabledOptical2022}%
  \BibitemOpen
  \bibfield  {author} {\bibinfo {author} {\bibfnamefont {Z.}~\bibnamefont
  {Wang}}, \bibinfo {author} {\bibfnamefont {Y.~C.}\ \bibnamefont {Lin}},
  \bibinfo {author} {\bibfnamefont {K.}~\bibnamefont {Zhang}}, \bibinfo
  {author} {\bibfnamefont {W.}~\bibnamefont {Wu}},\ and\ \bibinfo {author}
  {\bibfnamefont {S.}~\bibnamefont {Huang}},\ }\href
  {https://doi.org/10.48550/arXiv.2210.05630} {\bibinfo {title}
  {{{EllipsoNet}}: {{Deep-learning-enabled}} optical ellipsometry for complex
  thin films}} (\bibinfo {year} {2022}),\ \Eprint
  {https://arxiv.org/abs/2210.05630} {arxiv:2210.05630 [physics]} \BibitemShut
  {NoStop}%
\bibitem [{\citenamefont {Luce}\ \emph {et~al.}(2023)\citenamefont {Luce},
  \citenamefont {Mahdavi}, \citenamefont {Wankerl},\ and\ \citenamefont
  {Marquardt}}]{luceInvestigationInverseDesign2023}%
  \BibitemOpen
  \bibfield  {author} {\bibinfo {author} {\bibfnamefont {A.}~\bibnamefont
  {Luce}}, \bibinfo {author} {\bibfnamefont {A.}~\bibnamefont {Mahdavi}},
  \bibinfo {author} {\bibfnamefont {H.}~\bibnamefont {Wankerl}},\ and\ \bibinfo
  {author} {\bibfnamefont {F.}~\bibnamefont {Marquardt}},\ }\href
  {https://doi.org/10.1088/2632-2153/acb48d} {\bibfield  {journal} {\bibinfo
  {journal} {Machine Learning: Science and Technology}\ }\textbf {\bibinfo
  {volume} {4}},\ \bibinfo {pages} {015014} (\bibinfo {year}
  {2023})}\BibitemShut {NoStop}%
\bibitem [{\citenamefont {Ma}\ \emph {et~al.}(2023)\citenamefont {Ma},
  \citenamefont {Wang},\ and\ \citenamefont
  {Guo}}]{maOptoGPTFoundationModel2023}%
  \BibitemOpen
  \bibfield  {author} {\bibinfo {author} {\bibfnamefont {T.}~\bibnamefont
  {Ma}}, \bibinfo {author} {\bibfnamefont {H.}~\bibnamefont {Wang}},\ and\
  \bibinfo {author} {\bibfnamefont {L.~J.}\ \bibnamefont {Guo}},\ }\href
  {https://doi.org/10.48550/arXiv.2304.10294} {\bibinfo {title} {{{OptoGPT}}:
  {{A Foundation Model}} for {{Inverse Design}} in {{Optical Multilayer Thin
  Film Structures}}}} (\bibinfo {year} {2023}),\ \Eprint
  {https://arxiv.org/abs/2304.10294} {arxiv:2304.10294 [physics]} \BibitemShut
  {NoStop}%
\bibitem [{\citenamefont {Jiang}\ and\ \citenamefont
  {Fan}(2020)}]{jiangMultiobjectiveCategoricalGlobal2020}%
  \BibitemOpen
  \bibfield  {author} {\bibinfo {author} {\bibfnamefont {J.}~\bibnamefont
  {Jiang}}\ and\ \bibinfo {author} {\bibfnamefont {J.~A.}\ \bibnamefont
  {Fan}},\ }\href {https://doi.org/10.1515/nanoph-2020-0407} {\bibfield
  {journal} {\bibinfo  {journal} {Nanophotonics}\ }\textbf {\bibinfo {volume}
  {10}},\ \bibinfo {pages} {361} (\bibinfo {year} {2020})}\BibitemShut
  {NoStop}%
\bibitem [{\citenamefont {{Shiyuan Liu}}\ \emph {et~al.}(2023)\citenamefont
  {{Shiyuan Liu}}, \citenamefont {{Xiuguo Chen}},\ and\ \citenamefont {{Shuo
  Liu}}}]{shiyuanliuSmartEllipsometryPhysicsinformed2023}%
  \BibitemOpen
  \bibfield  {author} {\bibinfo {author} {\bibnamefont {{Shiyuan Liu}}},
  \bibinfo {author} {\bibnamefont {{Xiuguo Chen}}},\ and\ \bibinfo {author}
  {\bibnamefont {{Shuo Liu}}},\ }\href
  {https://doi.org/10.21203/rs.3.rs-3205511/v1} {\bibinfo {title} {Smart
  ellipsometry with physics-informed deep learning}} (\bibinfo {year} {2023}),\
  \Eprint {https://arxiv.org/abs/rs.3.rs-3205511} {Research
  Square:rs.3.rs-3205511} \BibitemShut {NoStop}%
\bibitem [{\citenamefont {Moreau}(2023)}]{moreau_pymoosh_2023}%
  \BibitemOpen
  \bibfield  {author} {\bibinfo {author} {\bibfnamefont {A.}~\bibnamefont
  {Moreau}},\ }\href@noop {} {\bibinfo {title} {Pymoosh}},\ \bibinfo
  {howpublished} {https://github.com/AnMoreau/PyMoosh} (\bibinfo {year}
  {2023})\BibitemShut {NoStop}%
\bibitem [{\citenamefont {Langevin}\ \emph {et~al.}(2023)\citenamefont
  {Langevin}, \citenamefont {Bennet}, \citenamefont {Wiecha}, \citenamefont
  {Chevalier}, \citenamefont {Teytaud},\ and\ \citenamefont
  {Moreau}}]{pymoosh}%
  \BibitemOpen
  \bibfield  {author} {\bibinfo {author} {\bibfnamefont {D.}~\bibnamefont
  {Langevin}}, \bibinfo {author} {\bibfnamefont {P.}~\bibnamefont {Bennet}},
  \bibinfo {author} {\bibfnamefont {P.}~\bibnamefont {Wiecha}}, \bibinfo
  {author} {\bibfnamefont {P.}~\bibnamefont {Chevalier}}, \bibinfo {author}
  {\bibfnamefont {O.}~\bibnamefont {Teytaud}},\ and\ \bibinfo {author}
  {\bibfnamefont {A.}~\bibnamefont {Moreau}},\ }\href
  {https://doi.org/10.5281/zenodo.10261964} {\bibinfo {title}
  {Anmoreau/pymoosh: Resub' release}} (\bibinfo {year} {2023})\BibitemShut
  {NoStop}%
\bibitem [{\citenamefont {Defrance}\ \emph {et~al.}(2016)\citenamefont
  {Defrance}, \citenamefont {Lema{\^i}tre}, \citenamefont {Ajib}, \citenamefont
  {Benedicto}, \citenamefont {Mallet}, \citenamefont {Poll{\`e}s},
  \citenamefont {Plumey}, \citenamefont {Mihailovic}, \citenamefont {Centeno},
  \citenamefont {Cirac{\`i}}, \citenamefont {Smith},\ and\ \citenamefont
  {Moreau}}]{defranceMooshNumericalSwiss2016}%
  \BibitemOpen
  \bibfield  {author} {\bibinfo {author} {\bibfnamefont {J.}~\bibnamefont
  {Defrance}}, \bibinfo {author} {\bibfnamefont {C.}~\bibnamefont
  {Lema{\^i}tre}}, \bibinfo {author} {\bibfnamefont {R.}~\bibnamefont {Ajib}},
  \bibinfo {author} {\bibfnamefont {J.}~\bibnamefont {Benedicto}}, \bibinfo
  {author} {\bibfnamefont {E.}~\bibnamefont {Mallet}}, \bibinfo {author}
  {\bibfnamefont {R.}~\bibnamefont {Poll{\`e}s}}, \bibinfo {author}
  {\bibfnamefont {J.-P.}\ \bibnamefont {Plumey}}, \bibinfo {author}
  {\bibfnamefont {M.}~\bibnamefont {Mihailovic}}, \bibinfo {author}
  {\bibfnamefont {E.}~\bibnamefont {Centeno}}, \bibinfo {author} {\bibfnamefont
  {C.}~\bibnamefont {Cirac{\`i}}}, \bibinfo {author} {\bibfnamefont
  {D.}~\bibnamefont {Smith}},\ and\ \bibinfo {author} {\bibfnamefont
  {A.}~\bibnamefont {Moreau}},\ }\href {https://doi.org/10.5334/jors.100}
  {\bibfield  {journal} {\bibinfo  {journal} {Journal of Open Research
  Software}\ }\textbf {\bibinfo {volume} {4}},\ \bibinfo {pages} {e13}
  (\bibinfo {year} {2016})}\BibitemShut {NoStop}%
\bibitem [{\citenamefont {Randles}\ \emph {et~al.}(2017)\citenamefont
  {Randles}, \citenamefont {Pasquetto}, \citenamefont {Golshan},\ and\
  \citenamefont {Borgman}}]{jupyter}%
  \BibitemOpen
  \bibfield  {author} {\bibinfo {author} {\bibfnamefont {B.~M.}\ \bibnamefont
  {Randles}}, \bibinfo {author} {\bibfnamefont {I.~V.}\ \bibnamefont
  {Pasquetto}}, \bibinfo {author} {\bibfnamefont {M.~S.}\ \bibnamefont
  {Golshan}},\ and\ \bibinfo {author} {\bibfnamefont {C.~L.}\ \bibnamefont
  {Borgman}},\ }in\ \href@noop {} {\emph {\bibinfo {booktitle} {2017 ACM/IEEE
  Joint Conference on Digital Libraries (JCDL)}}}\ (\bibinfo {organization}
  {IEEE},\ \bibinfo {year} {2017})\ pp.\ \bibinfo {pages} {1--2}\BibitemShut
  {NoStop}%
\bibitem [{\citenamefont {Giessen}\ and\ \citenamefont
  {Vogelgesang}(2009)}]{giessenGlimpsingWeakMagnetic2009}%
  \BibitemOpen
  \bibfield  {author} {\bibinfo {author} {\bibfnamefont {H.}~\bibnamefont
  {Giessen}}\ and\ \bibinfo {author} {\bibfnamefont {R.}~\bibnamefont
  {Vogelgesang}},\ }\href {https://doi.org/10.1126/science.1181552} {\bibfield
  {journal} {\bibinfo  {journal} {Science}\ }\textbf {\bibinfo {volume}
  {326}},\ \bibinfo {pages} {529} (\bibinfo {year} {2009})}\BibitemShut
  {NoStop}%
\bibitem [{\citenamefont {Lalanne}\ and\ \citenamefont
  {Morris}(1996)}]{lalanne_highly_1996}%
  \BibitemOpen
  \bibfield  {author} {\bibinfo {author} {\bibfnamefont {P.}~\bibnamefont
  {Lalanne}}\ and\ \bibinfo {author} {\bibfnamefont {G.~M.}\ \bibnamefont
  {Morris}},\ }\href@noop {} {\bibfield  {journal} {\bibinfo  {journal} {JOSA
  A}\ }\textbf {\bibinfo {volume} {13}} (\bibinfo {year} {1996})}\BibitemShut
  {NoStop}%
\bibitem [{\citenamefont {Granet}\ and\ \citenamefont
  {Guizal}(1996)}]{granet_efficient_1996}%
  \BibitemOpen
  \bibfield  {author} {\bibinfo {author} {\bibfnamefont {G.}~\bibnamefont
  {Granet}}\ and\ \bibinfo {author} {\bibfnamefont {B.}~\bibnamefont
  {Guizal}},\ }\bibfield  {journal} {\bibinfo  {journal} {JOSA A}\ }\textbf
  {\bibinfo {volume} {13}},\ \href {https://doi.org/10.1364/JOSAA.13.001019}
  {10.1364/JOSAA.13.001019} (\bibinfo {year} {1996})\BibitemShut {NoStop}%
\bibitem [{\citenamefont {Hughes}(1995)}]{hughes1995multiscale}%
  \BibitemOpen
  \bibfield  {author} {\bibinfo {author} {\bibfnamefont {T.~J.}\ \bibnamefont
  {Hughes}},\ }\href@noop {} {\bibfield  {journal} {\bibinfo  {journal}
  {Computer methods in applied mechanics and engineering}\ }\textbf {\bibinfo
  {volume} {127}},\ \bibinfo {pages} {387} (\bibinfo {year}
  {1995})}\BibitemShut {NoStop}%
\bibitem [{\citenamefont {Muller}\ \emph {et~al.}(2018)\citenamefont {Muller},
  \citenamefont {Brisebarre}, \citenamefont {De~Dinechin}, \citenamefont
  {Jeannerod}, \citenamefont {Lefevre}, \citenamefont {Melquiond},
  \citenamefont {Revol}, \citenamefont {Stehl{\'e}}, \citenamefont {Torres}
  \emph {et~al.}}]{muller2018handbook}%
  \BibitemOpen
  \bibfield  {author} {\bibinfo {author} {\bibfnamefont {J.-M.}\ \bibnamefont
  {Muller}}, \bibinfo {author} {\bibfnamefont {N.}~\bibnamefont {Brisebarre}},
  \bibinfo {author} {\bibfnamefont {F.}~\bibnamefont {De~Dinechin}}, \bibinfo
  {author} {\bibfnamefont {C.-P.}\ \bibnamefont {Jeannerod}}, \bibinfo {author}
  {\bibfnamefont {V.}~\bibnamefont {Lefevre}}, \bibinfo {author} {\bibfnamefont
  {G.}~\bibnamefont {Melquiond}}, \bibinfo {author} {\bibfnamefont
  {N.}~\bibnamefont {Revol}}, \bibinfo {author} {\bibfnamefont
  {D.}~\bibnamefont {Stehl{\'e}}}, \bibinfo {author} {\bibfnamefont
  {S.}~\bibnamefont {Torres}}, \emph {et~al.},\ }\href@noop {} {\emph {\bibinfo
  {title} {Handbook of floating-point arithmetic}}}\ (\bibinfo  {publisher}
  {Springer},\ \bibinfo {year} {2018})\BibitemShut {NoStop}%
\bibitem [{\citenamefont {Solnyshkov}\ \emph {et~al.}(2021)\citenamefont
  {Solnyshkov}, \citenamefont {Malpuech}, \citenamefont {St-Jean},
  \citenamefont {Ravets}, \citenamefont {Bloch},\ and\ \citenamefont
  {Amo}}]{solnyshkov2021microcavity}%
  \BibitemOpen
  \bibfield  {author} {\bibinfo {author} {\bibfnamefont {D.~D.}\ \bibnamefont
  {Solnyshkov}}, \bibinfo {author} {\bibfnamefont {G.}~\bibnamefont
  {Malpuech}}, \bibinfo {author} {\bibfnamefont {P.}~\bibnamefont {St-Jean}},
  \bibinfo {author} {\bibfnamefont {S.}~\bibnamefont {Ravets}}, \bibinfo
  {author} {\bibfnamefont {J.}~\bibnamefont {Bloch}},\ and\ \bibinfo {author}
  {\bibfnamefont {A.}~\bibnamefont {Amo}},\ }\href@noop {} {\bibfield
  {journal} {\bibinfo  {journal} {Optical Materials Express}\ }\textbf
  {\bibinfo {volume} {11}},\ \bibinfo {pages} {1119} (\bibinfo {year}
  {2021})}\BibitemShut {NoStop}%
\bibitem [{\citenamefont {Kretschmann}\ and\ \citenamefont
  {Raether}(1968)}]{kretschmannNotizenRadiativeDecay1968}%
  \BibitemOpen
  \bibfield  {author} {\bibinfo {author} {\bibfnamefont {E.}~\bibnamefont
  {Kretschmann}}\ and\ \bibinfo {author} {\bibfnamefont {H.}~\bibnamefont
  {Raether}},\ }\href {https://doi.org/10.1515/zna-1968-1247} {\bibfield
  {journal} {\bibinfo  {journal} {Zeitschrift f\"ur Naturforschung A}\ }\textbf
  {\bibinfo {volume} {23}},\ \bibinfo {pages} {2135} (\bibinfo {year}
  {1968})}\BibitemShut {NoStop}%
\bibitem [{\citenamefont {Pendry}(2000)}]{pendry_negative_2000}%
  \BibitemOpen
  \bibfield  {author} {\bibinfo {author} {\bibfnamefont {J.~B.}\ \bibnamefont
  {Pendry}},\ }\href {https://doi.org/10.1103/PhysRevLett.85.3966} {\bibfield
  {journal} {\bibinfo  {journal} {Physical Review Letters}\ }\textbf {\bibinfo
  {volume} {85}},\ \bibinfo {pages} {3966} (\bibinfo {year} {2000})},\ \bibinfo
  {note} {publisher: American Physical Society}\BibitemShut {NoStop}%
\bibitem [{\citenamefont {Santbergen}\ \emph {et~al.}(2010)\citenamefont
  {Santbergen}, \citenamefont {Goud}, \citenamefont {Zeman}, \citenamefont {van
  Roosmalen},\ and\ \citenamefont {van Zolingen}}]{santbergen2010am1}%
  \BibitemOpen
  \bibfield  {author} {\bibinfo {author} {\bibfnamefont {R.}~\bibnamefont
  {Santbergen}}, \bibinfo {author} {\bibfnamefont {J.}~\bibnamefont {Goud}},
  \bibinfo {author} {\bibfnamefont {M.}~\bibnamefont {Zeman}}, \bibinfo
  {author} {\bibfnamefont {J.}~\bibnamefont {van Roosmalen}},\ and\ \bibinfo
  {author} {\bibfnamefont {R.~C.}\ \bibnamefont {van Zolingen}},\ }\href@noop
  {} {\bibfield  {journal} {\bibinfo  {journal} {Solar energy materials and
  solar cells}\ }\textbf {\bibinfo {volume} {94}},\ \bibinfo {pages} {715}
  (\bibinfo {year} {2010})}\BibitemShut {NoStop}%
\bibitem [{\citenamefont {Tamir}(1986)}]{tamir1986nonspecular}%
  \BibitemOpen
  \bibfield  {author} {\bibinfo {author} {\bibfnamefont {T.}~\bibnamefont
  {Tamir}},\ }\href@noop {} {\bibfield  {journal} {\bibinfo  {journal} {JOSA
  A}\ }\textbf {\bibinfo {volume} {3}},\ \bibinfo {pages} {558} (\bibinfo
  {year} {1986})}\BibitemShut {NoStop}%
\bibitem [{\citenamefont {Polles}\ \emph {et~al.}(2010)\citenamefont {Polles},
  \citenamefont {Moreau},\ and\ \citenamefont {Granet}}]{polles2010light}%
  \BibitemOpen
  \bibfield  {author} {\bibinfo {author} {\bibfnamefont {R.}~\bibnamefont
  {Polles}}, \bibinfo {author} {\bibfnamefont {A.}~\bibnamefont {Moreau}},\
  and\ \bibinfo {author} {\bibfnamefont {G.}~\bibnamefont {Granet}},\
  }\href@noop {} {\bibfield  {journal} {\bibinfo  {journal} {Optics letters}\
  }\textbf {\bibinfo {volume} {35}},\ \bibinfo {pages} {3237} (\bibinfo {year}
  {2010})}\BibitemShut {NoStop}%
\bibitem [{\citenamefont {Petit}(1989)}]{petit1989ondes}%
  \BibitemOpen
  \bibfield  {author} {\bibinfo {author} {\bibfnamefont {R.}~\bibnamefont
  {Petit}},\ }\href@noop {} {\bibfield  {journal} {\bibinfo  {journal} {(No
  Title)}\ } (\bibinfo {year} {1989})}\BibitemShut {NoStop}%
\bibitem [{\citenamefont {Khaywah}\ \emph {et~al.}(2021)\citenamefont
  {Khaywah}, \citenamefont {Potdevin}, \citenamefont {R{\'e}veret},
  \citenamefont {Mahiou}, \citenamefont {Ouerdane}, \citenamefont {D{\'e}sert},
  \citenamefont {Parola}, \citenamefont {Chadeyron}, \citenamefont {Centeno},
  \citenamefont {Smaali} \emph {et~al.}}]{khaywah2021large}%
  \BibitemOpen
  \bibfield  {author} {\bibinfo {author} {\bibfnamefont {M.}~\bibnamefont
  {Khaywah}}, \bibinfo {author} {\bibfnamefont {A.}~\bibnamefont {Potdevin}},
  \bibinfo {author} {\bibfnamefont {F.}~\bibnamefont {R{\'e}veret}}, \bibinfo
  {author} {\bibfnamefont {R.}~\bibnamefont {Mahiou}}, \bibinfo {author}
  {\bibfnamefont {Y.}~\bibnamefont {Ouerdane}}, \bibinfo {author}
  {\bibfnamefont {A.}~\bibnamefont {D{\'e}sert}}, \bibinfo {author}
  {\bibfnamefont {S.}~\bibnamefont {Parola}}, \bibinfo {author} {\bibfnamefont
  {G.}~\bibnamefont {Chadeyron}}, \bibinfo {author} {\bibfnamefont
  {E.}~\bibnamefont {Centeno}}, \bibinfo {author} {\bibfnamefont
  {R.}~\bibnamefont {Smaali}}, \emph {et~al.},\ }\href@noop {} {\bibfield
  {journal} {\bibinfo  {journal} {The Journal of Physical Chemistry C}\
  }\textbf {\bibinfo {volume} {125}},\ \bibinfo {pages} {7780} (\bibinfo {year}
  {2021})}\BibitemShut {NoStop}%
\bibitem [{\citenamefont {Smith}\ \emph {et~al.}(1992)\citenamefont {Smith},
  \citenamefont {Houde-Walter},\ and\ \citenamefont {Forbes}}]{smith1992mode}%
  \BibitemOpen
  \bibfield  {author} {\bibinfo {author} {\bibfnamefont {R.~E.}\ \bibnamefont
  {Smith}}, \bibinfo {author} {\bibfnamefont {S.}~\bibnamefont
  {Houde-Walter}},\ and\ \bibinfo {author} {\bibfnamefont {G.}~\bibnamefont
  {Forbes}},\ }\href@noop {} {\bibfield  {journal} {\bibinfo  {journal} {IEEE
  journal of quantum electronics}\ }\textbf {\bibinfo {volume} {28}},\ \bibinfo
  {pages} {1520} (\bibinfo {year} {1992})}\BibitemShut {NoStop}%
\bibitem [{\citenamefont {Valiev}\ \emph {et~al.}(2018)\citenamefont {Valiev},
  \citenamefont {Vasilescu},\ and\ \citenamefont {Herbsleb}}]{pypi}%
  \BibitemOpen
  \bibfield  {author} {\bibinfo {author} {\bibfnamefont {M.}~\bibnamefont
  {Valiev}}, \bibinfo {author} {\bibfnamefont {B.}~\bibnamefont {Vasilescu}},\
  and\ \bibinfo {author} {\bibfnamefont {J.}~\bibnamefont {Herbsleb}},\ }in\
  \href@noop {} {\emph {\bibinfo {booktitle} {Proceedings of the 2018 26th ACM
  Joint Meeting on European Software Engineering Conference and Symposium on
  the Foundations of Software Engineering}}}\ (\bibinfo {year} {2018})\ pp.\
  \bibinfo {pages} {644--655}\BibitemShut {NoStop}%
\bibitem [{\citenamefont {Cosentino}\ \emph {et~al.}(2016)\citenamefont
  {Cosentino}, \citenamefont {Luis},\ and\ \citenamefont {Cabot}}]{github}%
  \BibitemOpen
  \bibfield  {author} {\bibinfo {author} {\bibfnamefont {V.}~\bibnamefont
  {Cosentino}}, \bibinfo {author} {\bibfnamefont {J.}~\bibnamefont {Luis}},\
  and\ \bibinfo {author} {\bibfnamefont {J.}~\bibnamefont {Cabot}},\ }in\
  \href@noop {} {\emph {\bibinfo {booktitle} {Proceedings of the 13th
  International Conference on Mining Software Repositories}}}\ (\bibinfo {year}
  {2016})\ pp.\ \bibinfo {pages} {137--141}\BibitemShut {NoStop}%
\bibitem [{\citenamefont {Tanhua}\ \emph {et~al.}(2019)\citenamefont {Tanhua},
  \citenamefont {Pouliquen}, \citenamefont {Hausman}, \citenamefont
  {O’brien}, \citenamefont {Bricher}, \citenamefont {De~Bruin}, \citenamefont
  {Buck}, \citenamefont {Burger}, \citenamefont {Carval}, \citenamefont {Casey}
  \emph {et~al.}}]{fairdata}%
  \BibitemOpen
  \bibfield  {author} {\bibinfo {author} {\bibfnamefont {T.}~\bibnamefont
  {Tanhua}}, \bibinfo {author} {\bibfnamefont {S.}~\bibnamefont {Pouliquen}},
  \bibinfo {author} {\bibfnamefont {J.}~\bibnamefont {Hausman}}, \bibinfo
  {author} {\bibfnamefont {K.}~\bibnamefont {O’brien}}, \bibinfo {author}
  {\bibfnamefont {P.}~\bibnamefont {Bricher}}, \bibinfo {author} {\bibfnamefont
  {T.}~\bibnamefont {De~Bruin}}, \bibinfo {author} {\bibfnamefont {J.~J.}\
  \bibnamefont {Buck}}, \bibinfo {author} {\bibfnamefont {E.~F.}\ \bibnamefont
  {Burger}}, \bibinfo {author} {\bibfnamefont {T.}~\bibnamefont {Carval}},
  \bibinfo {author} {\bibfnamefont {K.~S.}\ \bibnamefont {Casey}}, \emph
  {et~al.},\ }\href@noop {} {\bibfield  {journal} {\bibinfo  {journal}
  {Frontiers in Marine Science}\ }\textbf {\bibinfo {volume} {6}},\ \bibinfo
  {pages} {440} (\bibinfo {year} {2019})}\BibitemShut {NoStop}%
\bibitem [{\citenamefont {Khaireh-Walieh}\ \emph {et~al.}(2023)\citenamefont
  {Khaireh-Walieh}, \citenamefont {Langevin}, \citenamefont {Bennet},
  \citenamefont {Teytaud}, \citenamefont {Moreau},\ and\ \citenamefont
  {Wiecha}}]{khaireh2023newcomer}%
  \BibitemOpen
  \bibfield  {author} {\bibinfo {author} {\bibfnamefont {A.}~\bibnamefont
  {Khaireh-Walieh}}, \bibinfo {author} {\bibfnamefont {D.}~\bibnamefont
  {Langevin}}, \bibinfo {author} {\bibfnamefont {P.}~\bibnamefont {Bennet}},
  \bibinfo {author} {\bibfnamefont {O.}~\bibnamefont {Teytaud}}, \bibinfo
  {author} {\bibfnamefont {A.}~\bibnamefont {Moreau}},\ and\ \bibinfo {author}
  {\bibfnamefont {P.~R.}\ \bibnamefont {Wiecha}},\ }\href@noop {} {\bibfield
  {journal} {\bibinfo  {journal} {Nanophotonics}\ } (\bibinfo {year}
  {2023})}\BibitemShut {NoStop}%
\bibitem [{\citenamefont {Wiecha}(2023)}]{wiechaNewcomerGuideDeep2023}%
  \BibitemOpen
  \bibfield  {author} {\bibinfo {author} {\bibfnamefont {P.~R.}\ \bibnamefont
  {Wiecha}},\ }\href@noop {} {\bibinfo {title} {A newcomer's guide to deep
  learning for inverse design in nano-photonics}},\ \bibinfo {howpublished}
  {https://gitlab.com/wiechapeter/newcomer\_guide\_dl\_inversedesign} (\bibinfo
  {year} {2023})\BibitemShut {NoStop}%
\bibitem [{\citenamefont {Bennet}\ \emph {et~al.}(2023)\citenamefont {Bennet},
  \citenamefont {Langevin}, \citenamefont {Essoual}, \citenamefont
  {Khaireh-Walieh}, \citenamefont {Teytaud}, \citenamefont {Wiecha},\ and\
  \citenamefont {Moreau}}]{bennet2023illustrated}%
  \BibitemOpen
  \bibfield  {author} {\bibinfo {author} {\bibfnamefont {P.}~\bibnamefont
  {Bennet}}, \bibinfo {author} {\bibfnamefont {D.}~\bibnamefont {Langevin}},
  \bibinfo {author} {\bibfnamefont {C.}~\bibnamefont {Essoual}}, \bibinfo
  {author} {\bibfnamefont {A.}~\bibnamefont {Khaireh-Walieh}}, \bibinfo
  {author} {\bibfnamefont {O.}~\bibnamefont {Teytaud}}, \bibinfo {author}
  {\bibfnamefont {P.}~\bibnamefont {Wiecha}},\ and\ \bibinfo {author}
  {\bibfnamefont {A.}~\bibnamefont {Moreau}},\ }\href@noop {} {\bibinfo {title}
  {An illustrated tutorial on global optimization in nanophotonics}} (\bibinfo
  {year} {2023}),\ \Eprint {https://arxiv.org/abs/2309.09760} {arXiv:2309.09760
  [physics.optics]} \BibitemShut {NoStop}%
\bibitem [{\citenamefont {Melnyk}\ and\ \citenamefont
  {Harrison}(1970)}]{melnyk1970theory}%
  \BibitemOpen
  \bibfield  {author} {\bibinfo {author} {\bibfnamefont {A.~R.}\ \bibnamefont
  {Melnyk}}\ and\ \bibinfo {author} {\bibfnamefont {M.~J.}\ \bibnamefont
  {Harrison}},\ }\href@noop {} {\bibfield  {journal} {\bibinfo  {journal}
  {Physical Review B}\ }\textbf {\bibinfo {volume} {2}},\ \bibinfo {pages}
  {835} (\bibinfo {year} {1970})}\BibitemShut {NoStop}%
\bibitem [{\citenamefont {Benedicto}\ \emph {et~al.}(2015)\citenamefont
  {Benedicto}, \citenamefont {Polles}, \citenamefont {Cirac{\`\i}},
  \citenamefont {Centeno}, \citenamefont {Smith},\ and\ \citenamefont
  {Moreau}}]{benedicto2015numerical}%
  \BibitemOpen
  \bibfield  {author} {\bibinfo {author} {\bibfnamefont {J.}~\bibnamefont
  {Benedicto}}, \bibinfo {author} {\bibfnamefont {R.}~\bibnamefont {Polles}},
  \bibinfo {author} {\bibfnamefont {C.}~\bibnamefont {Cirac{\`\i}}}, \bibinfo
  {author} {\bibfnamefont {E.}~\bibnamefont {Centeno}}, \bibinfo {author}
  {\bibfnamefont {D.~R.}\ \bibnamefont {Smith}},\ and\ \bibinfo {author}
  {\bibfnamefont {A.}~\bibnamefont {Moreau}},\ }\href@noop {} {\bibfield
  {journal} {\bibinfo  {journal} {JOSA A}\ }\textbf {\bibinfo {volume} {32}},\
  \bibinfo {pages} {1581} (\bibinfo {year} {2015})}\BibitemShut {NoStop}%
\end{thebibliography}%

\end{document}